\documentclass[10pt,a4paper]{article}
\usepackage{mathrsfs}
\usepackage{fancyhdr}
\usepackage[top=30truemm,bottom=30truemm,left=20truemm,right=50truemm]{geometry}
\usepackage{indentfirst}
\usepackage{authblk}
\usepackage{graphicx}% Include figure files
\usepackage{bm,color}% bold math   \textcolor{red} ON
\renewenvironment{abstract}{\begin{quotation}{\normalsize }}{\end{quotation}}
\makeatletter
\newcommand{\email}[1]{\newcommand{\@email}{E-mail: #1}}
\renewcommand{\maketitle}{
\newpage\null
    \vspace{2em}
 {\LARGE\bfseries\noindent\ignorespaces\@title\par}
 \vspace{1em}%
 {\large\noindent\ignorespaces\@author\par}
 \vspace{2mm}
 {\normalsize\noindent\ignorespaces\@email\par}
\vspace{1em}
}
\renewcommand{\Affilfont}{\normalsize}
\renewcommand\@author{\ifx\AB@affillist\AB@empty\AB@author\else
      \ifnum\value{affil}>\value{Maxaffil}\def\rlap##1{##1}%
     \vspace{1mm} \AB@authlist\\\AB@affillist
    \else  \AB@authors\fi\fi}
\makeatother
\usepackage{amsmath,amssymb}
\usepackage{braket}
\title{Cluster states in stable and unstable nuclei}
\author[1]{M. Kimura}
\affil[1]{Department of Physics, Hokkaido University, Sapporo 060-0810, Japan}
\email{masaaki@nucl.sci.hokudai.ac.jp}
\begin{document}
\maketitle

%%%abstract
\begin{abstract}
{\bf Abstract:}
 In this contribution, I will discuss two topics related to the clustering of atomic nuclei. The
 first is dual character of the ground state. Recently, it was pointed out that the ground
 states of atomic nuclei have dual character of shell and cluster implying that both of the
 single-particle excitation and cluster excitation occur as the fundamental excitation modes. 
 The isoscalar monopole and dipole transitions are regarded as the triggers to induce the cluster
 excitation. By referring the Hartree-Fock and antisymmetrized molecular dynamics calculations,
 the dual character and those  transitions are explained. The second is the linear-chain state
 of $^{16}{\rm C}$ in which the linearly aligned 3$\alpha$ particles are sustained by the valence
 neutrons. The rather convincing evidences for this exotic state are very recently reported by
 several experiments. Here, I introduce a couple of theoretical analysis and predictions.
\end{abstract}
%%%%%%%

%%%keyword
{\bf keywords:} Clustering of atomic nuclei, Unstable nuclei
%%%%%%

\section*{Introduction}
The study of the clustering of atomic nuclei has long history its very beginning \cite{gam31} is even
earlier than the finding of neutron \cite{cha32}. In 1950's, K. Wildermuth and his coworkers
\cite{wil58,wil59} revived a microscopic cluster model called resonating group method \cite{whe37}
and greatly developed nuclear cluster study enabling quantitative descriptions of nuclear
structure and reactions \cite{wil77}. In 1960's, D. Brink  revived another microscopic cluster
model called Bloch-Brink model \cite{bri66} today. Both models have been playing important role as
vital driving force to promote the studies of clustering of various nuclei
\cite{bec10,bec12,bec14}, but at the same time, those models have been criticized. Bayman and Bohr
\cite{bay58} showed that those cluster model wave functions become identical to the Elliott
$SU(3)$ shell model wave function \cite{ell58} at the limit where the inter-cluster distance
becomes zero, which is called Bayman-Bohr theorem. From this fact, it was argued that the states
described by a cluster model wave function can be also described by the shell model, and hence,
the cluster models were not describing a new state but just rewriting an ordinary shell
model state. This argument was invalidated by the discovery of the well developed cluster states
in  $^{16}{\rm O}$ and $^{20}{\rm Ne}$ \cite{hor68} in which the inter-cluster distances are large 
and cannot be described by the limited number of the shell model wave functions. A good example of
this is the Hoyle  state (the $0^+_2$ state) of $^{12}{\rm C}$ which has a dilute gas-like
3$\alpha$ cluster structure \cite{ueg77,kam81,des87,kan98,toh01,nef04}. It was shown that the
shell model needs a huge model space larger than 50$\hbar\omega$ excitation \cite{yam08-1} to 
describe the Hoyle state, which is not manageable by any modern computers. On the other hand, a
single Tohsaki-Horiuchi-Schuck-Ro\"pke (THSR) wave function which is a modern cluster model can
describe  this state surprisingly well \cite{toh01,yam08-1,fun06,fun15}.  

It is interesting to know that Bayman-Bohr theorem is interpreted in a different meaning today, and 
is regarded as a strong evidence for the clustering in atomic nuclei \cite{suz87,suz89,yam08}. The
mathematical equivalence of the shell model and cluster model wave function is interpreted as 
``{\it dual character of shell and cluster}'' \cite{bec10,hor10}.  Namely, this equivalence
means that the  degrees-of-freedom of cluster excitation is embedded even in a pure shell model
ground state. Therefore, it implies that both of the single-particle and cluster excitations 
are possible from the ground state and indicates that the clustering is one of the fundamental
degrees-of-freedom of nuclear excitations.  Indeed, it was recently found that isoscalar (IS)
monopole \cite{yam08} and dipole \cite{kim16} excitations do activate the degrees-of-freedom of
cluster excitation and generate pronounced cluster states. This is a point what I want to
discuss in the first half of this contribution.    

Another point I discuss in this contribution is exotic clustering of neutron-rich nuclei. 
From the early stage of the study of neutron-rich nuclei, it was realized that the excess neutrons
can induce or stabilize exotic cluster structure which cannot appear in ordinary stable nuclei. 
A good example is Be isotopes \cite{sey81,oer96,ita00,des02,eny95,eny03}. $^{8}{\rm Be}$ has an
extreme 2$\alpha$ cluster structure, but is unstable against $\alpha$ decay. If we add one or two
neutrons to this system, the cluster structure is stabilized and the system ($^{9}{\rm Be}$ and
$^{10}{\rm Be}$ ) is bound. If we continue to add excess neutrons, it enhances the
clustering. $^{11}{\rm Be}$ and  $^{12}{\rm Be}$ are known to have pronounced 2$\alpha$ cluster
core surrounded by valence neutrons and neutron magic number $N=8$ is broken in those isotopes. It 
was found that because of the two center nature of the 2$\alpha$ cluster core, a special class of
neutron single-particle orbits different from the ordinary spherical shell is formed in Be
isotopes. This so-called ``molecular-orbits'' \cite{sey81,oer96,ita00} play an important role for
the  enhancement of the clustering and the breaking of the neutron magic number. This finding
motivated the search for very exotic cluster structure of linear-chain state in neutron-rich
Carbon isotopes which is composed of  linearly aligned three $\alpha$ particles with surrounding
neutrons \cite{oer97,oer06,ita01}. In these days, many theoretical studies have been performed 
\cite{mar10,suh11,kim14,zha15,bab16} and very promising experimental data are reported
\cite{fre14,koy15,fri15,yam15,Tia16,del16}. I'll report the present status of the research
briefly, and more detailed discussion can be found in the contribution by N. Itagaki. 

This contribution is organized as follows. In the next section, I explain the dual character of
shell and cluster referring the study of $^{8}{\rm Be}$, $^{16}{\rm O}$ and $^{20}{\rm Ne}$ made
by Maruhn {\it et al.}. Then I discuss the IS monopole and dipole excitations in $^{20}{\rm Ne}$
to show that those excitations activate the degrees-of-freedom of cluster excitation embedded in
the ground state. After the discussion of stable nuclei, I discuss the linear-chain in Cabon
isotopes, in particular, I focus on the linear-chain state in  $^{16}{\rm C}$. Finally, I
summarize this contribution. 

 \section*{Dual character of shell and cluster; an analysis of the Hartree-Fock ground states
 of light nuclei}

In this section, we discuss the dual character of shell and cluster. 
It is known that the ground state wave functions of $^{8}{\rm Be}$, $^{16}{\rm O}$ and 
$^{20}{\rm Ne}$ obtained by the shell model or mean-field models have large overlap with the
cluster model wave function.  This is due to the mathematical equivalence of the shell model and
cluster model wave functions at small inter-cluster distance as proved by Bayman-Bohr
theorem \cite{bay58}. Therefore, this equivalence does not necessarily mean that the ground state
has cluster structure. 

However, it is important to realize that this equivalence  means that the degrees-of-freedom of
cluster excitation is embedded in the ground state even if it has an ideal shell
structure \cite{suz87,suz89,yam08}. Therefore, it implies that both of the single-particle and
cluster excitations are possible from the ground state. We shall call this nature of the ground
state  ``dual character of shell and cluster''. Indeed, in the next section, we will show that the
IS monopole and dipole transitions actually activate the degrees-of-freedom of cluster excitation
and generates pronounced cluster states.   

To understand this dual character, an analysis of the Hartree-Fock wave function made by Maruhn
{\it et al.} \cite{mar06} is suggestive and provides a very good insight. Therefore, the discussion 
made in this section is mainly based on his work.

   \subsection*{Pronounced 2$\bm \alpha$ clustering in $^{\bm 8}{\bf Be}$}
   $^{8}{\rm Be}$ is a metastable nucleus with pronounced 2$\alpha$ cluster structure. By the Green
   Function Monte Carlo (GFMC) calculation \cite{wir00}, the inter-cluster distance is estimated
   approximately 4 fm, which is larger than the inter-cluster distance of two touching $\alpha$
   particles. Therefore, $^{8}{\rm Be}$ is regarded as an extreme case of the clustering
   and  a good starting point for the discussion. 

   I first introduce the Bloch-Brink wave function \cite{bri66} for $^{8}{\rm Be}$ that is
   composed of    2$\alpha$ clusters with the inter-cluster distance $R$,
   \begin{align}
    &\Phi_{BB}(R) = n_0\mathcal A'\left\{
    \psi_{\alpha}\Bigl(-\frac{\bm R}{2}\Bigr) \psi_{\alpha}\Bigl(\frac{\bm R}{2}\Bigr)
    \right\},\quad     \bm R = (0,0,R).
   \end{align}
    Here, $\psi_{\alpha}(\bm R)$ is the wave function of $\alpha$ cluster placed at the position 
   $\bm R$ and represented by the harmonic oscillator wave function,
   \begin{align}
    &\psi_{\alpha}(\bm R) = \mathcal{A}
    \Set{\phi_1(\bm r_1;\bm R)\phi_2(\bm r_2;\bm R)\phi_3(\bm r_3;\bm R)\phi_4(\bm r_4;\bm R)},
    \label{eq:BB1}\\
    &\phi_i(\bm r;\bm R) = \left(\frac{1}{\pi\sigma^2}\right)^{3/4}
    \exp\Set{-\frac{1}{2\sigma^2}(\bm r - \bm R)^2}\chi_i,\quad
    \chi_i =  n\uparrow, n\downarrow, p\uparrow \text{or } p\downarrow.
   \end{align}
   It is known that Bloch-Brink wave function can be rewritten into the following form \cite{hor77}, 
   \begin{align}
    &\Phi_{BB}(R) = \phi_{ cm}(\bm r_{cm})\cdot n_0 \mathcal A'
    \set{\chi_{BB}(\bm r)\phi_{\alpha}\phi_{\alpha}},\label{eq:BB2}\\
    &\chi_{BB}(\bm r) 
    =\sum_{Nl}(-)^{(N-l)/2}\sqrt{\frac{(2l+1)N!}{(N-l)!!(N+l+1)!!}}
    \frac{(R^2/(2\sigma'^{2}))^{N/2}}{\sqrt{N!}}e^{-R^2/(4\sigma'^2)} 
    \mathcal R_{Nl0}(\bm r), \label{eq:BB3}\\
    &\phi_{cm}(\bm r_{cm}) = \left(\frac{A}{\pi\sigma^2}\right)^{3/4}e^{-Ar_{cm}^2/(2\sigma^2) }.
    \nonumber
   \end{align}
   Here $\phi_{\alpha}$ is the internal wave function of $\alpha$ particle, {\it i.e.} the
   center-of-mass coordinate of $\alpha$ particle is removed from $\psi_\alpha$ given in
   Eq. (\ref{eq:BB1}). $\phi_{cm}(\bm r_{cm})$ represents the center-of-mass wave function of 
   the total system. The wave function of the inter-cluster motion $\chi_{BB}(\bm r)$ is expanded 
   by the harmonic oscillator wave functions, 
   $\mathcal R_{Nlm}(\bm r)\equiv R_{Nl}(r)Y_{lm}(\hat r)$ whose oscillator width is scaled by 
   reduced mass $\sigma'=\sqrt{8/(4\cdot 4)}\sigma=\sigma/\sqrt{2}$. $N$ represents the principal quantum number of the
   inter-cluster  motion. 

   From Eqs. (\ref{eq:BB2}) and (\ref{eq:BB3}),   we observe the following properties of
   Bloch-Brink wave function, which we utilize  for the discussions below.
   \begin{enumerate}
    \item The principal quantum number $N$ must be equal to or larger than the lowest Pauli
	  allowed value $N_0=4$. This is proved as follows. The total principal quantum number of
	  the shell model wave function for $^{8}{\rm Be}$  must be equal to or larger than 4,
	  because at least four nucleons must occupy $0p$ orbit. On the other hand, the principal
	  quantum number of r.h.s. of Eq. (\ref{eq:BB2}) is equal to that of $\chi_{BB}(\bm r)$
	  given in Eq. (\ref{eq:BB3}), because the quantum number of $\phi_\alpha$ is zero. Hence,
	  the condition $N\geq 4$ must be satisfied, and the summation over $N$ runs for 
	  $N\geq N_0=4$, otherwise  the r.h.s. of Eq. (\ref{eq:BB2}) identically vanishes. 
    \item At the limit of $R\rightarrow 0$, Eq. (\ref{eq:BB2}) becomes identical to the $SU(3)$
	  shell model wave function belongs to the irreducible representation of 
	  $(\lambda,\mu)=(4,0)$, which is so-called Bayman-Bohr theorem. If we expand 
	  Eq. (\ref{eq:BB3}) and $N^{-1}(R)=\sqrt{\braket{\Phi_{BB}(R)|\Phi_{BB}(R)}}$ in a power 
	  series of $R$, their leading terms are proportional to $R^{N_0}$. Therefore, at the
	  limit of $R\rightarrow 0$, the normalized wave function $N(R)\Phi_{BB}(R)$ becomes
	  identical to the wave function having the quantum number $(N_x,N_y,N_z)=(0,0,N_0)$
	  which is nothing but the the irreducible representation of $(\lambda,\mu)=(4,0)$.
    \item As the inter-cluster distance $R$ increases, the system has, of course, prominent
	  cluster structure. This means that the inter-cluster wave function with large principal
	  quantum number $N$ is superposed coherently as observed from Eq. (\ref{eq:BB3}).
   \end{enumerate}
   Keeping the above properties in mind, we refer to the analysis of the Hartree-Fock ground
   state made in Ref. \cite{mar06}. The overlap between the single Bloch-Brink wave function and the
   Hartree-Fock ground state is defined as,
   \begin{align}
    \mathcal{O}_{BB} = \frac{|\braket{\Phi_{BB}(R)|\Psi_{HF}}|^2}
    {\braket{\Phi_{BB}(R)|\Phi_{BB}(R)}},\label{eq:obb}
   \end{align}
   where the inter-cluster distance $R$ and the size of the Gaussian wave packet $\sigma$ are so
   chosen to maximize the overlap $\mathcal{O}_{BB}$. The projector to the Bloch-Brink wave
   function is also introduced, 
   \begin{align}
    P_{BB} = \sum_{ij}\ket{\Phi_{BB}(R_i)}B^{-1}_{ij}\bra{\Phi_{BB}(R_j)},
   \end{align}
   where the summation over $i$ and $j$ runs for the discretized set of the inter-cluster distance 
   $R_i$, $i=1,2,...,N$. $B^{-1}$ is the inverse of the overlap matrix 
   $B_{ij}=\braket{\Phi_{BB}(R_i)|\Phi_{BB}(R_j)}$. With this projector, the amount of the cluster
   component in the Hartree-Fock ground state which we denote by $\mathcal{O}_{GCM}$ is also 
   evaluated,
   \begin{align}
    \mathcal{O}_{GCM} = \braket{\Psi_{HF}|P_{BB}|\Psi_{HF}}. \label{eq:ogcm}
   \end{align}
   \begin{table}[h]
    \begin{center}
     \caption{The overlap between the Hartree-Fock ground state of $^{8}{\rm Be}$ and 
     Bloch-Brink wave function. The inter-cluster distance $R$ and the size of the Gaussian wave
     packet $\sigma$ are so chosen to maximize the overlap. $\sigma_\alpha$ is the size of the
     Gaussian wave packet which maximizes the overlap between Eq. (\ref{eq:BB1}) and Hartree-Fock
     wave function for $^{4}{\rm He}$. This table is reconstructed from the data given in
     Ref. \cite{mar06}.}\label{tab:Be8}
     \begin{tabular}{cccccccc}
      \hline\hline
      &&\multicolumn{3}{c}{Bloch-Brink}&&\multicolumn{2}{c}{GCM}\\
      \cline{3-5}\cline{7-8}
      Skyrme&$\sigma_\alpha({\rm fm})$&$\mathcal{O}_{BB} (\%)$& $R ({\rm fm})$ & 
      $\sigma({\rm fm})$&& $\mathcal{O}_{GCM} (\%)$&$\sigma({\rm fm})$  \\
      \hline
      SkI3 & 1.67 & 82 & 2.70 & 1.68 && 98 & 1.65 \\
      SkI4 & 1.68 & 81 & 2.64 & 1.69 && 97 & 1.64 \\
      Sly6 & 1.72 & 81 & 2.68 & 1.73 && 97 & 1.68 \\
      SkM* & 1.61 & 71 & 2.20 & 1.68 && 82 & 1.62 \\
      \hline\hline
     \end{tabular}
    \end{center}
   \end{table}
    Table \ref{tab:Be8} shows these overlaps  obtained for $^{8}{\rm Be}$ by using several Skyrme
    parameter sets. The results in the table indicate that the Hartree-Fock ground states have
    large overlap of $\mathcal{O}_{BB}$, and hence, are well approximated with single Bloch-Brink
    wave function with proper choice of $R$ and $\sigma$. It is also remarkable the radius
    parameters $\sigma$ are very close to  those for the free $\alpha$ particle. The distance $R$
    shows two $\alpha$ particles are well separated, although it is not as large as that of the
    GFMC result \cite{wir00}. It is noted that if the angular momentum projection was performed, the
    optimal distance became as large as 4 fm, because the strongly deformed configuration gains
    larger binding energy by the projection. When we allow the superposition of multiple
    Bloch-Brink wave functions, the overlap further increases and amounts to almost 100\%
    except for the result of SkM*. Those analysis assert following points.  
    \begin{enumerate}
     \item The Hartree-Fock is based on the independent particle picture, while the Bloch-Brink 
	   suggests quite different picture in which four nucleons are strongly correlated to form
	   $\alpha$ clusters. Nevertheless, both wave functions are almost identical as shown in
	   Tab. \ref{tab:Be8}. This means Hartree-Fock wave function is capable to describe
	   cluster correlation in the ground state.  
     \item The estimated inter-cluster distance  clearly deviates from the $SU(3)$ shell model
	   limit of $R\rightarrow 0$ offering prominent 2$\alpha$ clustering in accordance with
	   the GFMC result. From Eq. (\ref{eq:BB3}), we see that large fraction of the wave
	   function with large $N$ is contained in the ground state. 
    \end{enumerate}
   
   \subsection*{Tetrahedral configuration of 4$\bm \alpha$ particles in $^{\bm 16}{\bf O}$}
   The ground state of $^{16}{\rm O}$ is, of course, well represented by the $(0s)^4(0p)^{12}$
   closed shell configuration and the cluster correlation is much less important than 
   $^{8}{\rm Be}$. Hence, $^{16}{\rm O}$ is regarded as an extreme opposite case to  $^{8}{\rm
   Be}$.  In Ref. \cite{mar06}, all the Skyrme parameter sets yielded almost identical spherical
   ground state.   The result of the overlap analysis is listed in Table \ref{tab:O16}. 
   \begin{table}[h]
    \begin{center}
     \caption{The overlap between the Hartree-Fock ground state of $^{16}{\rm O}$ and 
     Bloch-Brink wave function of four $\alpha$ particles with tetrahedral configuration. 
     The length of the side of tetrahedron is 0.01 fm. The size of the Gaussian wave packet
     $\sigma$ are so chosen to maximize the overlap. This table is reconstructed from the data
     given in Ref. \cite{mar06}.}\label{tab:O16}   
     \begin{tabular}{ccc}
      \hline\hline
      Skyrme&$\mathcal{O}_{BB} (\%)$& $\sigma({\rm fm})$\\
      \hline
      SkI3 & 96 & 1.76 \\
      SkI4 & 96 & 1.76 \\
      Sly6 & 96 & 1.79 \\
      SkM* & 96 & 1.78 \\
      \hline\hline
     \end{tabular}
    \end{center}
   \end{table}
   It was found, as expected, that the tetrahedral configuration of four $\alpha$ particles has
   quite large overlap with the ground state. It does not mean the enhanced clustering of the
   ground state, because the size of the tetrahedron is quite small as its sides are only 0.01 fm,
   but means that $\alpha$ cluster wave function becomes mathematically identical to the
   shell model wave function at the limit of the inter-cluster distance $R\rightarrow 0$ owing to
   the antisymmetrization (Bayman-Bohr theorem). However, it is important to note that this
   equivalence of shell model and  cluster model wave function plays very important role as
   emphasized in Ref. \cite{yam08}. From this equivalence, the shell model wave function of the
   ground state is rewritten by using the cluster model wave function,  
   \begin{align}
    \Phi(0^+_1) &= N_0\mathcal{A}\Set{\mathcal R_{400}(\bm r)
    \phi_{^{12}{\rm C}}(0^+_1)\phi_{\alpha}}\\
    &= N_2\mathcal{A}\Set{[\mathcal R_{420}(\bm r)\otimes
    \phi_{^{12}{\rm C}}(2^+_1)]_{00}\phi_{\alpha}}\\
    &= N_4\mathcal{A}\Set{[\mathcal R_{440}(\bm r)\otimes
    \phi_{^{12}{\rm C}}(4^+_1)]_{00}\phi_{\alpha}}.
   \end{align}
   Those expressions imply that the degrees-of-freedom of cluster excitation is embedded even in
   an ideal shell model state. Because if the wave function of the inter-cluster motion
   $\mathcal{R}_{Nlm}(\bm r)$ is excited, it will yield  pronounced cluster states. It was found
   that  the IS monopole excitation is a strong trigger to excite the inter-cluster motion, and
   hence a very good probe for clustering \cite{yam08} which we shall discuss in this contribution.

   \subsection*{Transient nature of $^{\bm 20}{\bf Ne}$ and $\bm \alpha$+$^{\bf 16}{\bf O}$
   clustering }  
   We have seen two extreme cases, $^{8}{\rm Be}$ with extreme 2$\alpha$ clustering and 
   $^{16}{\rm O}$  with an almost ideal shell model ground state. $^{20}{\rm Ne}$ discussed here
   has transient nature between them. Namely, the $\alpha$+$^{16}{\rm O}$ cluster structure of
   the ground state is well established. At the same time, it is also known that the distortion of 
   the clusters is important because of the spin-orbit interaction and the formation of the mean
   field.     
   \begin{figure}[h]
    \begin{center}
     \includegraphics[width=0.6\hsize]{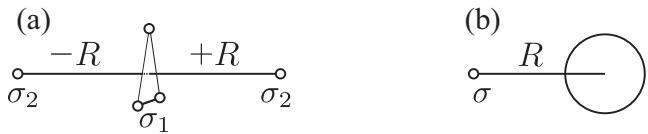}
     \caption{(a) Optimum 5$\alpha$ configuration that maximizes the overlap with the
     Hartree-Fock ground state of $^{20}{\rm Ne}$. A small triangle of 3$\alpha$ particles with
     radius $\sigma_1$ locates at the center and perpendicular to the longest axis of
     nucleus. Along this axis, two $\alpha$ particles with radius $\sigma_2$ are positioned at
     $\pm R$. (b) $\alpha$+$^{16}{\rm O}$ configuration which also has large overlap with the
     Hartree-Fock ground state. $\alpha$ particle and Hartree-Fock ground state of $^{16}{\rm O}$
     are placed with the distance $R$. This figure is reconstructed from
     Ref. \cite{mar06}}\label{fig:Ne20conf}  
    \end{center}
   \end{figure}
   \begin{table}[h]
    \begin{center}
     \caption{The overlap between the Hartree-Fock ground state of $^{20}{\rm Ne}$ and 
     5$\alpha$ Bloch-Brink wave function. The parameters of Bloch-Brink wave function are
     explained in the text and Fig. \ref{fig:Ne20conf}. This table is reconstructed from the data
     given in Ref. \cite{mar06}.}\label{tab:Ne20-1}    
     \begin{tabular}{ccccc}
      \hline\hline
      Skyrme&$\mathcal{O}_{BB} (\%)$& $R ({\rm fm})$ & 
      $\sigma_1({\rm fm})$&$\sigma_2({\rm fm})$  \\
      \hline
      SkI3 & 53 & 1.91 & 1.78 & 1.71  \\
      SkI4 & 49 & 1.86 & 1.78 & 1.71  \\
      Sly6 & 47 & 1.84 & 1.80 & 1.74  \\
      SkM* & 36 & 1.59 & 1.77 & 1.73  \\
      \hline\hline
     \end{tabular}
    \end{center}
   \end{table}

   The obtained Hartree-Fock ground state shows a strong prolate deformation. For this ground
   state, it was found that two different cluster configurations have large overlap. The first one
   is 5$\alpha$ configuration shown in the Fig. \ref{fig:Ne20conf} (a). A small triangle of
   3$\alpha$  particles with radius $\sigma_1$ locates at the center and perpendicular to the
   longest axis of nucleus. Along this axis, two $\alpha$ particles with radius $\sigma_2$ are
   positioned at $\pm R$. This configuration is consistent with the result reported in
   Ref. \cite{zha94}. The 
   optimal values of those parameters are listed in Tab. \ref{tab:Ne20-1}. The distance $R$ is
   around 1.9 fm and two $\alpha$ particles are well separated from the triangle implying
   non-negligible $\alpha$ correlation in the ground state, but it is not as large as that of 
   $^{8}{\rm Be}$. The maximum value of the overlap    is around 50\% and much less than the case
   of $^{8}{\rm Be}$. Those results indicate the reduction of the cluster correlation compared to
   $^{8}{\rm Be}$ and increased importance of the cluster distortion effect due to the spin-orbit
   interaction and the formation of mean field. 
   
   Another configuration shown in Fig. \ref{fig:Ne20conf} (b) also has large overlap and is more 
   suggestive. In this configuration, $\alpha$ particle and Hartree-Fock  wave function
   of $^{16}{\rm O}$ are placed with the distance $R$ and projected to the positive parity,
   \begin{align}
    &\Phi_{BB}(R) = n_0\mathcal A'\left\{
    \psi_{\alpha}\Bigl(-\frac{4}{5}\bm R\Bigr) \psi_{^{16}{\rm O}}\Bigl(\frac{1}{5}\bm R\Bigr)
    \right\},\quad     \bm R = (0,0,R), \\
    &\Phi_{BB}^+(R) = \frac{1+P_x}{2} \Phi_{BB}(R),
   \end{align}
   where $P_x$ denotes parity operator. The size of $\alpha$ particle and inter cluster distance
   $R$ are again optimized to maximize the overlap between $\Phi_{BB}^+(R)$ and Hartree-Fock
   ground state, as listed in Tab. \ref{tab:Ne20-2}. 
   \begin{table}[h]
    \begin{center}
     \caption{The overlap between the Hartree-Fock ground state of $^{20}{\rm Ne}$ and 
     $\alpha$+$^{16}{\rm O}$ Bloch-Brink wave function projected to positive parity. 
     This table is reconstructed from the data given in Ref. \cite{mar06}.}\label{tab:Ne20-2} 
     \begin{tabular}{ccccccc}
      \hline\hline
      &\multicolumn{3}{c}{Bloch-Brink}&&\multicolumn{2}{c}{GCM}\\
      \cline{2-4}\cline{6-7}
      Skyrme&$\mathcal{O}_{BB} (\%)$& $R ({\rm fm})$ & 
      $\sigma({\rm fm})$&& $\mathcal{O}_{GCM} (\%)$&$\sigma({\rm fm})$  \\
      \hline
      SkI3 & 52 & 2.7 & 1.59 && 54 & 1.56 \\
      SkI4 & 48 & 2.5 & 1.62 && 49 & 1.52 \\
      Sly6 & 48 & 2.6 & 1.66 && 49 & 1.55 \\
      SkM* & 36 & 2.4 & 1.64 && 36 & 1.53 \\
      \hline\hline
     \end{tabular}
    \end{center}
   \end{table}
   It is interesting to note that this configuration has almost the same  overlap with the
   5$\alpha$ configuration. Furthermore, in this case, the size of $\alpha$ particle is close to
   that of free $\alpha$ particle and the inter-cluster distance is comparable with that of
   $^{8}{\rm Be}$. As we show later, the inter-cluster distance further increases when the
   angular momentum projection is performed. If we allow superposition of the Bloch-Brink wave
   functions, the overlap slightly increases.
   
   Those results suggest that the ground state of $^{20}{\rm Ne}$ can be written in the following
   form, 
   \begin{align}
    \Phi(0^+_1) =  \sum_{N=N_0}^\infty e_N n_N\mathcal{A}
    \Set{\mathcal{R}_{N00}(\bm r)\phi_{\alpha}\phi_{^{16}{\rm O}}}
    +\Phi_{\perp}, \quad N_0=8
   \end{align}
   Here $n_N \mathcal{A}\Set{\mathcal{R}_{N00}(\bm r)\phi_{\alpha}\phi_{^{16}{\rm O}}}$ is a
   normalized cluster model wave function in which the inter-cluster motion is described by the
   harmonic oscillator wave function $\mathcal{R}_{N00}(\bm r)$, and 
   $\Phi_{\perp}$ is non-cluster wave function that is orthogonal to 
   $\mathcal{A}\Set{\mathcal{R}_{N00}(\bm r)\phi_{\alpha}\phi_{^{16}{\rm O}}}$.
   From Tab. \ref{tab:Ne20-2}, the sum of the squared coefficient of superposition and the norm of
   $\Phi_\perp$ should be around 0.5,  
   \begin{align}
    \sum_{N=N_0}^\infty |e_N|^2 \simeq 0.5, \quad \braket{\Phi_\perp|\Phi_\perp} \simeq 0.5.
   \end{align}
   By a similar consideration to the case of $^{8}{\rm Be}$ and $^{16}{\rm O}$, we regard
   that the degrees-of-freedom of cluster excitation is embedded in the ground state of
   $^{20}{\rm Ne}$. If the inter-cluster motion $\mathcal{R}_{Nlm}(\bm r)$ is excited, it will
   populates pronounced $\alpha$+$^{16}{\rm O}$ cluster states. 

   \subsection*{Dual character of shell and cluster}
   Here, we summarize the insights obtained by the cluster analysis of the Hartree-Fock ground
   states.  
   \begin{enumerate}
    \item All of nuclei examined here, $^{8}{\rm Be}$, $^{16}{\rm O}$ and $^{20}{\rm Ne}$, have
	  large overlap with Bloch-Brink wave functions. This does not necessarily mean the
	  prominent clustering, because at the limit of $R\rightarrow 0$, the cluster wave
	  function becomes identical to the shell model wave function (Bayman-Bohr theorem).
    \item However, the large overlap means that the ground state has the dual character of shell
	  and cluster. Namely, the ground state wave function can be rewritten as a sum of cluster 
	  and non-cluster wave functions,
	  \begin{align}
	       \Phi(0^+_1) =  \sum_{N=N_0}^\infty e_N n_N\mathcal{A}
	       \Set{\mathcal{R}_{N00}(\bm r)\phi_{C_1}\phi_{C_2}}
	       +\Phi_{\perp}. \label{eq:BBsum}
	  \end{align}
	  This expression implies that the degrees-of-freedom of cluster excitation is embedded in
	  the ground state. If the inter-cluster motion $\mathcal{R}_{N00}(\bm r)$ is excited, it
	  yields excited states with  pronounced clustering. 
    \item $^{8}{\rm Be}$ has large overlap with Bloch-Brink wave function. Therefore, 
	  the norm of $\Phi_{\perp}$ in Eq. (\ref{eq:BBsum}) is rather small. Furthermore, because
	  the inter-cluster distance is very large, the wave function  with large principal
	  quantum number $N$ is coherently superposed. 
    \item $^{16}{\rm O}$  has almost 100\% overlap with Bloch-Brink wave function,  and hence,
	  $\Phi_{\perp}$ is negligible. Because the inter-cluster distance $R$ is very close to 
	  0, the wave function with principal quantum number $N_0=4$ dominates. 
    \item $^{20}{\rm Ne}$ has approximately 50\% overlap with Bloch-Brink wave function, and 
	  hence, the norm of $\Phi_{\perp}$ is approximately 0.5. The optimum inter-cluster
	  distance $R$ is not small which means the moderate clustering in the ground state. 
	  In other words, $e_{N}$ for larger values of $N>N_0 = 8$ are not negligible. 
   \end{enumerate}

\section*{Isoscalar monopole, dipole transitions and cluster states}
In the previous section, we have seen that the Hartree-Fock ground states of $^{8}{\rm Be}$,
$^{16}{\rm O}$ and $^{20}{\rm Ne}$ have large overlap with the cluster model (Bloch-Brink) wave
functions and discussed that the ground states has dual character of shell and cluster. This fact 
implies that both of the single-particle and cluster excitations possibly occur, and if the
inter-cluster motion is excited, prominent cluster states will be generated.

Here, by using $^{20}{\rm Ne}$ as an example,  we show that the IS monopole and dipole transitions
indeed induce the inter-cluster excitation to generate prominent cluster states. I first explain
the nodal and angular excited cluster states of $^{20}{\rm Ne}$, and derive the analytical formula
for transition matrix. It is shown that the transition matrices from the ground state to those
exited cluster states are as large as Weisskopf estimates even if the ground  state is not cluster
state but a pure shell model state.  Hence, the IS monopole and dipole excitations are regarded as
good probe for cluster states.  

Using Bloch-Brink wave function, I also show that the transition matrix is more enhanced, 
if the ground state has cluster structure.  More realistic calculation by using antisymmetrized
molecular dynamics (AMD) is also presented.

\subsection*{Nodal and angular excited cluster states}
The ground state of $^{20}{\rm Ne}$ is written as,
\begin{align}
 \Phi(0^+_1) =  \sum_{N=N_0}^\infty e_N n_N\mathcal{A}
 \Set{\mathcal{R}_{N00}(\bm r)\phi_{\alpha}\phi_{^{16}{\rm O}}}
 +\Phi_{\perp}. \label{eq:gswf0}
\end{align}
As already discussed, this expression of the ground state wave function implies that the
degrees-of-freedom of cluster excitation is embedded in the ground state, and the prominent
cluster states are generated by the excitation of the inter-cluster motion. 

The $\alpha$+$^{16}{\rm O}$ cluster states in $^{20}{\rm Ne}$ have been studied in detail and are
established well
\cite{mat75,buc75, hee76, fuj79,kru92,duf94,bo13,zhou15}. Therefore,
we already know the corresponding excited states which are populated by the excitation of
inter-cluster  motion. For example, if the nodal quantum number $n$ is increased, it yields the
excited $0^+$ state described as, 
\begin{align}
 &\Phi(0^+) = \sum_{N=N_0+2}^{\infty} f_N n_{N}
 \mathcal A'\Set{\mathcal R_{N00}(\bm r)\phi_{\alpha}\phi_{^{16}{\rm O}}}.\label{eq:ex0wf}
\end{align}
Here, the principal quantum number $N$ is equal to or larger than $N_0+2$, because of the
relation $N=2n+l$. The $0^+_4$ state of $^{20}$Ne observed as a broad resonance around 8.7 MeV has
large $\alpha$ width comparable with Wigner limit and is attributed to this class of nodal excited
cluster state.  

Besides the nodal excitation, the angular excitation of the inter-cluster motion
is also possible. For example, the angular excitation with $\Delta l=1$ (combined
with the nodal excitation) yields the $1^-$ state,  
\begin{align}
 &\Phi(1^-) = \sum_{N=N_0+1}^{\infty}  g_n n_{N}
 \mathcal A'\Set{\mathcal R_{N10}(r)\phi_{\alpha}\phi_{^{16}{\rm O}}},\label{eq:ex1wf}
\end{align}
where the principal quantum number $N$ is equal to or larger than $N_0+1$ because the angular
momentum of the inter-cluster motion is increased to $l=1$. The $1^-_1$ state observed  at 5.8 MeV
 also has large $\alpha$ width and is attributed to this class of angular 
excited cluster state. The cluster system with parity asymmetric intrinsic structure 
must have negative-parity states as well as positive-parity states. Therefore, this $1^-_1$ state
is very important as the evidence for the asymmetric cluster structure of 
$\alpha$+$^{16}{\rm O}$ \cite{hor68}.

\subsection*{Analytical expressions for transition matrix}
Using the wave functions given in  Eqs. (\ref{eq:ex0wf}) and (\ref{eq:ex1wf}), we discuss
an analytic expression for the IS monopole and dipole transitions from the ground state to 
the nodal and angular excited cluster states. The IS monopole and dipole operators 
$\mathcal M^{ IS0}, \mathcal M_\mu^{ IS1}$,  their transition probabilities $B(IS0), B(IS1)$ 
and reduced matrix elements $M^{IS0}, M^{IS1}$ are   
\begin{align}
 &\mathcal M^{ IS0}=\sum_{i=1}^{A}(\bm r_i - \bm r_{\rm cm})^2,\\
 &\mathcal M_\mu^{ IS1}=\sum_{i=1}^{A}(\bm r_i - \bm r_{\rm cm})^2
 \mathcal Y_{1\mu}(\bm r_i-\bm r_{\rm cm}), \\
 &B(IS0;0^+_1 \rightarrow 0^+_4) = |M^{IS0}|^2,\quad
 M^{IS0}=\braket{0^+_4||\mathcal M^{IS0}||0^+_1}
 = \braket{0^+_4|\mathcal M^{IS0}|0^+_1}, \label{eq:redmat0}\\
 &B(IS1;0^+_1 \rightarrow 1^-_1) = |M^{IS1}|^2,\quad
 M^{IS1}=\braket{1^-_1||\mathcal M^{IS1}||0^+_1}
 = \sqrt{3}\braket{1^-_1,J_z|\mathcal M^{IS1}_{J_z}|0^+_1}, \label{eq:redmat1}
\end{align}
where $\bm r_i$ denotes the $i$th nucleon coordinate, while $\bm r_{\rm cm}$ is the
center-of-mass coordinate of the system. The solid spherical harmonics are
defined as $ \mathcal Y_{\lambda\mu} (\bm r)\equiv r^\lambda Y_{\lambda\mu}(\hat r)$.

To clarify the relationship between the monopole transition and clustering,  we rewrite 
$\mathcal M_\mu^{IS0}$ in terms of the internal coordinates $\bm \xi_i$ of each cluster and the
inter-cluster coordinate $\bm r$ defined as,
\begin{align}
 &\bm \xi_i \equiv\left\{
 \begin{array}{l}
 \bm r_i - \bm R_{\alpha},\quad i\in \alpha\\
 \bm r_i - \bm R_{^{16}{\rm O}},\quad i\in {}^{16}{\rm O} ,\\
 \end{array}
\right.\quad
 \bm r \equiv \bm R_{\alpha} - \bm R_{^{16}{\rm O}}, \\
 &\bm R_{\alpha} \equiv \frac{1}{4}\sum_{i\in \alpha}\bm r_i,\quad
 \bm R_{^{16}{\rm O}} \equiv \frac{1}{16}\sum_{i\in ^{16}{\rm O}}\bm r_i,
\end{align}
where the center-of-mass of $\alpha$ and $^{16}{\rm O}$ clusters $\bm R_{\alpha}$ and 
$\bm R_{^{16}{\rm O}}$ are introduced. With these coordinates, $\mathcal M_\mu^{IS0}$ is 
rewritten as follows,
\begin{align}
 \mathcal{M}_\mu^{IS0} =& \sum_{i=1}^{A}(\bm r_i - \bm r_{\rm cm})^2 
 =\sum_{i\in \alpha}\Bigl(\bm \xi_i + \frac{4}{5}\bm r\Bigr)^2
 +\sum_{i\in ^{16}{\rm O}}\Bigl(\bm \xi_i - \frac{1}{5}\bm r\Bigr)^2 \nonumber\\
 =& \sum_{i\in \alpha}\xi_i^2  +\sum_{i\in ^{16}{\rm O}}\xi_i^2
 +\frac{16}{5}r^2
 \label{eq:opis0}
\end{align}
where the relations $\sum_{i\in \alpha} \bm \xi_i = \sum_{i\in ^{16}{\rm O}} \bm \xi_i = 0$ 
are used. This expression makes it clear that $\mathcal M^{\rm IS0}$ will populate nodal excited
$0^+$ cluster states, because if operated to the ground state wave function given in
Eq. (\ref{eq:gswf0}), the last term will induce the nodal excitation of the inter-cluster
motion. By a similar calculation, $\mathcal M^{IS1}_\mu$ is rewritten as follows
(see Ref. \cite{kim16} for the derivation),
\begin{align}
 \mathcal M^{IS1}_\mu =&  \sum_{i\in \alpha}\xi_i^2\mathcal Y_{1\mu}(\bm \xi_i)
 +\sum_{i\in ^{16}{\rm O}}\xi_i^2\mathcal Y_{1\mu}(\bm \xi_i)\nonumber\\
 &-\sqrt{\frac{32\pi}{9}}\Set{
 \frac{4}{5}\biggl[\sum_{i\in \alpha}\mathcal Y_{2}(\bm \xi_i)
 \otimes \mathcal Y_{1}(\bm r)\biggr]_{1\mu}
 -\frac{1}{5}\biggl[\sum_{i\in ^{16}{\rm O}}\mathcal Y_{2}(\bm \xi_i)
 \otimes \mathcal Y_{1}(\bm r)\biggr]_{1\mu}
 }\nonumber\\
 &+\left( \frac{4}{3}\sum_{i\in \alpha}\xi_i^2 
 -\frac{1}{3}\sum_{i\in ^{16}{\rm O}}\xi_i^2 \right)\mathcal Y_{1\mu}(\bm r) 
 +\frac{48}{25}r^2\mathcal Y_{1\mu}(\bm r). \label{eq:opis1}
\end{align}
From this expression, we find that $\mathcal M_\mu^{IS1}$ will populate $1^-$ cluster states
because  the terms depending on $\mathcal Y_{1\mu}(\bm r)$ and $r^2\mathcal Y_{1\mu}(\bm r)$ 
in the second and third lines will induce the  nodal and angular excitation of the inter-cluster
motion.  

If we assume that the ground state wave function is a pure shell model state, {\it i.e.} the
wave function of the inter-cluster motion is the harmonic oscillator wave function with the lowest
Pauli allowed  principal quantum number $N_0$,
\begin{align}
 &\Phi(0^+_1) = n_{N_0}
 \mathcal A'\Set{\mathcal R_{N_000}(\bm r)\phi_{\alpha}\phi_{^{16}{\rm O}}},\label{eq:gswf1}
\end{align}
then, it is possible to calculate the transition matrix analytically. By substituting
Eqs. (\ref{eq:ex0wf}), (\ref{eq:ex1wf})  and (\ref{eq:gswf1}) into Eqs. (\ref{eq:redmat0}) and
(\ref{eq:redmat1}), one gets
\begin{align}
 M^{IS0} =& f_{N_0+2}\sqrt{\frac{\mu_{N_0}}{\mu_{N_0+2}}}
 \braket{R_{N_00}|r^2|R_{N_0+20}} \label{eq:mat0}\\
 M^{IS1} =&
 \sqrt{\frac{3}{4\pi}}\Biggl[g_{N_0+1}\sqrt{\frac{\mu_{N_0}}{\mu_{N_0+1}}}
 \biggl\{
 \frac{16}{3}\left(\braket{r^2}_{\alpha} - \braket{r^2}_{^{16}{\rm O}}\right)
\braket{R_{N_00}|r|R_{N_0+11}}
 +\frac{48}{25}\braket{R_{N_00}|r^3|R_{N_0+11}}\biggr\} \nonumber\\
 &\qquad\qquad\qquad +\frac{3}{5}
 g_{N_0+3}\sqrt{\frac{\mu_{N_0}}{\mu_{N_0+3}}}\braket{R_{N_00}|r^3|R_{N_0+31}}\Biggr],
 \label{eq:mat1}
\end{align}
where $\braket{r^2}_{\alpha}$ and $\braket{r^2}_{^{16}{\rm O}}$ are the mean-square radius of
the clusters, and  the matrix elements of harmonic oscillator are given as,  
\begin{align}
 &\braket{R_{N_00}|r^2|R_{N_0+20}} = -\frac{\sigma'^2}{2}\sqrt{(N_0+2)(N_0+3)},\quad
 \braket{R_{N_00}|r|R_{N_0+11}} = \sigma'\sqrt{\frac{N_0+3}{2}},\nonumber\\
 &\braket{R_{N_00}|r^3|R_{N_0+11}} = \sigma'^3\frac{3N_0+5}{2}\sqrt{\frac{N_0+3}{2}},\nonumber\\
 &\braket{R_{N_00}|r^3|R_{N_0+31}} = -\sigma'^3\frac{\sqrt{(N_0+2)(N_0+5)}}{2}
 \sqrt{\frac{N_0+3}{2}}.  \label{eq:ho1}
\end{align}
And $\mu_N$ is so-called eigenvalue of RGM norm kernel defined as,
\begin{align}
 \mu_N = \braket{\mathcal{R}_{Nlm}(\bm r)\phi_{\alpha}\phi_{^{16}{\rm O}}|
 \mathcal A\set{\mathcal{R}_{Nlm}(\bm r)\phi_{\alpha}\phi_{^{16}{\rm O}}}},
\end{align}
which is analytically calculable. Detailed explanations of  above formula are given in
Refs. \cite{yam08,kim16}. 
\begin{table}[h]
 \begin{center}
\caption{List of the quantities used to evaluate Eq. (\ref{eq:mat0}) and (\ref{eq:mat1}). Radii of
  $\alpha$ and  $^{16}$O clusters are calculated from the measured charge radii given in 
 Ref. \cite{ang13}  and  listed in the units of   ${\rm fm}^2$. The oscillator parameters $\sigma$ 
 and $\sigma'$ are in  units of fm. Coefficients $f_N$ and $g_N$ are obtained by the AMD
  calculation explained in the next section.}\label{tab:value} 
  \begin{tabular}{ccccccc}
   \hline\hline
   $N_0$& $\mu_{N_0}$ & $\mu_{N_0+1}$&$\mu_{N_0+2}$ &$\mu_{N_0+3}$&
   $\braket{r^2}_{\alpha}$ &$\braket{r^2}_{^{16}{\rm O}}$\\
   \hline
    8 & 0.229 &  0.344 &0.510&0.620& $(1.46)^2$&$(2.57)^2$\\
   \hline
    $\sigma$ & $\sigma'$ &  $g_{N_0+1}$ & $f_{N_0+2}$&$g_{N_0+3}$& &\\
   \hline
    1.77 & 0.99 &$\sqrt{0.39}$ &$\sqrt{0.51}$&-$\sqrt{0.28}$  & &\\
   \hline\hline
  \end{tabular}
 \end{center}
\end{table}

To evaluate the magnitude of transition matrices, we adopt the values listed in Table
\ref{tab:value}, where the values of $\sigma$, $\sigma'$, $f_N$ and $g_N$ are taken from the AMD
result explained later. Other variables are analytically calculable or taken from the experimental
data.  Assignment of those values to Eqs. (\ref{eq:mat0}) and (\ref{eq:mat1}) yields the estimate
of  IS monopole transition,
\begin{align}
 M^{IS0} = -7.67f_{10} = -5.48\ \rm fm^2, \label{eq:es0}
\end{align}
and IS dipole transition,
\begin{align}
 M^{IS1} = 3.08 g_{9} -7.36 g_{11} = 5.82\ \rm fm^3. \label{eq:es1}
\end{align}

These results are compared with the single-particle estimates. Assuming the constant radial wave
function as usual \cite{mart}, Weisskopf estimates are given as 
\begin{align}
 &M_{\rm WU}^{IS0} = \frac{3}{5}(1.2A^{1/3})^2 \simeq 0.864 A^{2/3}
 \simeq 6.37\ \rm fm^2,\\
 &M_{\rm WU}^{IS1} = \sqrt{\frac{3}{4\pi}}\frac{3}{6}(1.2A^{1/3})^3 \simeq 0.422 A
 \simeq 8.44\ \rm fm^3,
\end{align}
which are slightly larger than but comparable with Eqs. (\ref{eq:es0}) and (\ref{eq:es1}).

Thus, the nodal and angular excited cluster states have strong IS monopole  and dipole transitions
from the ground state comparable with the Weisskopf estimate, even if the ground state is not a
cluster state but an ideal shell model state. Since the single-particle transition is usually
fragmented into many states, we expect that only the asymmetric cluster states can have strong
transition strengths. Furthermore, as we will show below, the IS dipole strength is further
amplified if the ground state has cluster structure.  

\subsection*{Numerical estimate by using Bloch-Brink wave function}
Here we show that the transition matrix $M^{IS1}$ is greatly amplified compared to the
estimates made in the previous subsection, if the ground state has cluster correlation.
I again employ Bloch-Brink wave function to discuss it,
\begin{align}
 &\Phi_{BB}(R) = \frac{1+\pi P_x}{2} n_0\mathcal A'\left\{
 \psi_{\alpha}\Bigl(-\frac{4}{5}\bm R\Bigr) \psi_{^{16}{\rm O}}\Bigl(\frac{1}{5}\bm R\Bigr)
 \right\},\label{eq:BB_Ne1}\quad \pi=\pm, \quad     \bm R = (0,0,R), 
\end{align}
and I project it to $0^+$ and $1^-$ states,
\begin{align}
 &\Phi^{l\pi}_{ BB}(R) = \frac{2l+1}{8\pi^2}\int d\Omega D^{l*}_{M0}(\Omega)
 R(\Omega)\Phi^\pi_{ BB}(R). 
\end{align}
Here $D^l_{MK}(\Omega)$ and $R(\Omega)$ denote Wigner $D$ function and rotation operator.
By using this wave function, I calculated the transition matrix,  
\begin{align}
 &M^{IS1}(R_{0},R_{1}) =
 \frac{\sqrt{3}
 \braket{\Phi_{\rm BB}^{1^-}(R_{1})|\mathcal M^{IS1}_0|\Phi_{\rm BB}^{0^+}(R_{0})}}
 {\sqrt{\braket{\Phi_{\rm BB}^{0^+}(R_{0})|\Phi_{\rm BB}^{0^+}(R_{0})}
 \braket{\Phi_{\rm BB}^{1^-}(R_{1})|\Phi_{\rm BB}^{1^-}(R_{1})}}}.
\end{align}
\begin{figure}[t]
 \begin{center}
  \includegraphics[width=0.5\hsize]{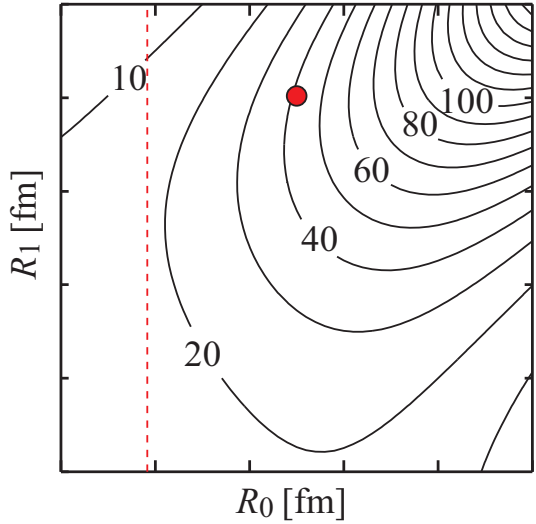}
  \caption{The transition matrix $M^{IS1}$  as function of the inter-cluster distances 
  in the ground state ($R_{0}$) and in the $1^-$ state ($R_{1}$). 
  The circle show the  approximate  position  of the  ground and the excited $1^-$ states obtained
  by AMD discussed in the next subsection. This figure is reconstructed from the data given in 
  Ref. \cite{kim16}}\label{fig:isd_brink}  
 \end{center}
\end{figure}
The calculated transition matrix $M^{IS1}$ is shown in Fig. \ref{fig:isd_brink} as
function of the inter-cluster distances $R_{0}$ of the ground state and
$R_{1}$ of the $1^-$ state. We can see that even for the small values of $R_{0}$ and $R_{1}$, 
$M^{IS1}$ is larger than the Weisskopf estimate. It is impressive that the 
matrix element is considerably amplified, as both of $R_{0}$ and $R_{1}$ growth. 

By more detailed calculation explained in the next section, the position of the ground and $1^-$
states are estimated approximately at the circle in Fig. \ref{fig:isd_brink}. Therefore, the
transition strength is considerably amplified and regarded as a good probe for asymmetric
clustering.

\subsection*{Realistic calculation with antisymmetrized molecular dynamics}
I have discussed that the IS monopole and dipole transitions are greatly enhanced for the nodal
and angular excited cluster states. However, in more realistic situation, there is the cluster
distortion effect as expressed by Eq. (\ref{eq:gswf0}). Therefore, for the quantitative
discussion, we need 
more realistic calculation which takes cluster distortion effect into account. For this purpose,
we introduce antisymmetrized molecular dynamics (AMD) which is a microscopic nuclear model able to 
describe both of cluster and non-cluster states.  First, we briefly explain the framework of
AMD. Readers are directed to  Refs. \cite{amd1,kim04,amd2}  for more detail. Then, we discuss the
numerical results obtained by AMD.  
\subsubsection*{Framework of antisymmetrized molecular dynamics}
In the present calculation, we employed the following microscopic $A$-body
Hamiltonian,  
\begin{align}
 {H} = \sum_{i=1}^A {t}(i) + \sum_{i<j}^A {v}_n(ij) + \sum_{i<j}^Z {v}_C(ij) - {t}_{\rm cm},
\end{align}
where the Gogny D1S interaction \cite{gog91} is used as an effective nucleon-nucleon interaction
$v_n$. Coulomb interaction $v_C$ is approximated by a sum of seven
Gaussians. The center-of-mass kinetic energy ${t}_{\rm cm}$ is exactly removed. 

The intrinsic wave function of AMD is a Slater determinant of nucleon wave packets 
$\varphi_i$ represented by a localized Gaussian, 
\begin{align}
 &\Phi_{\rm int} ={\mathcal A} \{\varphi_1\varphi_2...\varphi_A \},\\  \label{eq:wf_amd}
 &\varphi_i({\bm r}) = \exp\biggl\{-\sum_{\sigma=x,y,z}\nu_\sigma
  \Bigl(r_\sigma - 
  \frac{Z_{i\sigma}}{\sqrt{\nu_\sigma}}\Bigr)^2\biggr\}\chi_i\xi_i, \\
 &\chi_i = a_i\chi_\uparrow + b_i\chi_\downarrow,\quad
 \xi_i = {\rm proton} \ {\rm or} \ {\rm neutron},\nonumber
\end{align}
where $\chi_i$ and $\xi_i$ represent spin and isospin wave functions. The intrinsic wave function
is projected to the eigenstate of the parity, 
\begin{align}
 \Phi^\pi &= \frac{1+\pi P_x}{2}\Phi_{\rm int},\quad \pi = \pm.
  \label{EQ_INTRINSIC_WF}  
\end{align}
Then, the parameters of the intrinsic wave function, ${\bm Z}_i$, $a_i$, $b_i$ and $\nu_\sigma$,
are determined to minimize the expectation value of the Hamiltonian $\widetilde{E}$ that is
defined as   
\begin{align}
 \widetilde{E}&=\frac{\langle \Phi^\pi|\hat H|\Phi^\pi\rangle}{\langle
  \Phi^\pi|\Phi^\pi\rangle} + V_c,\quad
 V_c=v_\beta(\langle\beta\rangle-\beta)^2
  +v_\gamma(\langle\gamma\rangle-\gamma)^2.
\end{align}
Here the potential $V_c$  imposes the constraint on the quadrupole deformation of
intrinsic wave function parameterized by $\braket{\beta}$ and $\braket{\gamma}$ \cite{kim12}. 
The values of  $v_\beta$ and $v_\gamma$ are chosen large enough so that
$\braket{\beta},\braket{\gamma}$ are, after the energy minimization,  
equal to $\beta,\gamma$.  By the energy minimization, we obtain the optimized wave
function denoted by $\Phi^{\pi}_{\rm int}(\beta,\gamma)$ for given values of $(\beta, \gamma)$. 

After the energy minimization, we project out an eigenstate of angular momentum,
\begin{eqnarray}
 \Phi^{J\pi}_{MK}(\beta,\gamma) = \frac{2J+1}{8\pi^2}\int d\Omega
  D^{J*}_{MK}(\Omega)\hat{R}(\Omega)\Phi^{\pi}(\beta,\gamma).
\end{eqnarray} 
Then the wave functions with different values of quadrupole deformations $\beta$ and $\gamma$ are
superposed. 
\begin{align}
 &\Psi^{J\pi}_{Mp} = \sum_{Ki}g_{Kip}\Phi^{J\pi}_{MK}(\beta_i,\gamma_i),\label{eq:gcmwf}
\end{align}
and the coefficient of superposition $g_{Kip}$ and the eigenenergy $E^{J\pi}_p$ are determined 
by solving the Griffin-Hill-Wheeler equation \cite{hill53,hill57},
\begin{align}
 &\sum_{i'K'}{H^{J\pi}_{KiK'i'}g_{K'i'p}} = E^{J\pi}_p
 \sum_{i'K'}{N^{J\pi}_{KiK'i'}g_{K'i'p}},\\ 
 &H^{J\pi}_{KiK'i'} = \langle{\Phi^{J\pi}_{MKi}|\hat{H}|\Phi^{J\pi}_{MK'i'}}\rangle, \quad
 N^{J\pi}_{KiK'i'} = \langle{\Phi^{J\pi}_{MKi}|\Phi^{J\pi}_{MK'i'}}\rangle.
\end{align}
Using the wave function given in Eq. (\ref{eq:gcmwf}) the transition matrix element is calculated. 
It is noted that AMD is able to describe the distortion of clusters and the coupling between the
cluster states and non-cluster states, because all nucleons are treated independently.

\subsubsection*{Cluster states in $^{\bf 20}{\bf Ne}$ and IS monopole and dipole 
   transitions} 
\begin{figure}[h]
 \begin{center}
  \includegraphics[width=0.6\hsize]{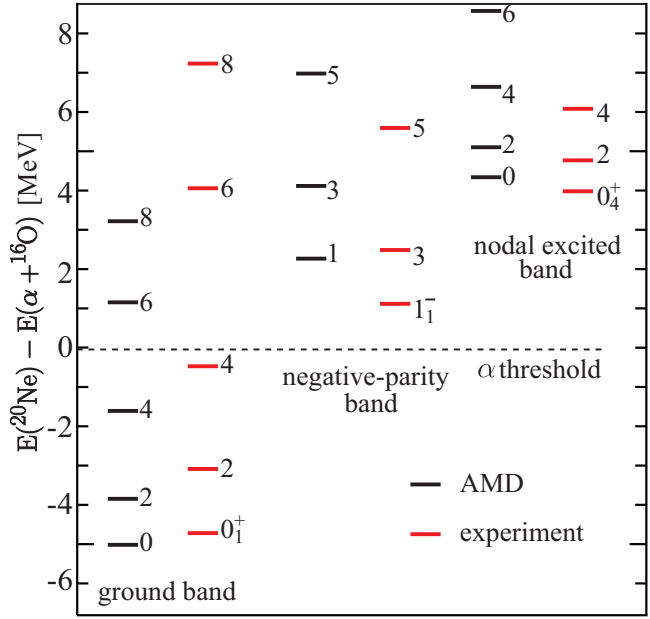}
  \caption{The observed and calculated $\alpha$+$^{16}{\rm O}$ cluster states in 
 $^{20}{\rm Ne}$ classified into three rotational bands. Energy is measured from the
 $\alpha$ threshold located at 4.7 MeV above the ground state.}\label{fig:Ne20level}   
 \end{center}
\end{figure}

The observed $\alpha$+$^{16}{\rm O}$ cluster bands are summarized in Fig. \ref{fig:Ne20level}
together with the result of AMD. The  $0^+_4$ state at 8.7 MeV (4 MeV above the $\alpha$
threshold) is identified as the nodal excited state described by the wave function given in 
Eq. (\ref{eq:ex0wf}). A rotational band is built on this state. The $1^-_1$ state at 5.8 MeV (1.1
MeV above the $\alpha$ threshold) is the angular excited cluster state described by
the wave function of Eq. (\ref{eq:ex1wf}). This $1^-_1$ state is very important because it is
regarded as the evidence for the asymmetric clustering with 
$\alpha$+$^{16}{\rm O}$ configuration.

Now, we examine the result of AMD. I calculated the amount of the cluster component defined by
Eq. (\ref{eq:obb}) and (\ref{eq:ogcm}) in each states, and identified the states with large
overlap as the $\alpha$+$^{16}{\rm O}$ cluster states. Those cluster states are shown in
Fig. \ref{fig:Ne20level} which reasonably agree with the observed spectrum. It is noted that AMD
calculation also reproduces non-cluster states which are not shown in Fig. \ref{fig:Ne20level} and 
readers are directed to Ref. \cite{kim04} for detailed discussions. Here, we focus on the ground
$0^+_1$, the nodal excited $0^+_4$ and the angular excited $1^-_1$ states, whose properties are
listed in Tab. \ref{tab:Ne20ism}.
\begin{table}[h]
 \begin{center}
  \caption{The properties of the ground, $0^+_4$ and $1^-_1$ states obtained by AMD. 
  The overlap with the Bloch-Brink wave functions are given in percentage and the estimated
  inter-cluster distance is in units of fm. $M^{IS0}$ and $M^{IS1}$ are the IS monopole and
  dipole transition matrix from the ground state to the $0^+_4$ and $1^-_1$ states given in units
  of $\rm fm^2$ and $\rm fm^3$, respectively. }\label{tab:Ne20ism} 
  \begin{tabular}{cccccccccccccc}
   \hline\hline
   \multicolumn{3}{c}{$0^+_1$}&&\multicolumn{3}{c}{$0^+_4$}&&\multicolumn{3}{c}{$1^-_1$}\\
   \cline{1-3}\cline{5-7}\cline{9-11}
   $\mathcal{O}_{BB}$&$R$&$\mathcal{O}_{GCM}$&&$\mathcal{O}_{BB}$&$R$&$\mathcal{O}_{GCM}$
   &&$\mathcal{O}_{BB}$&$R$&$\mathcal{O}_{GCM}$&$M^{IS0}$&$M^{IS1}$\\
   \hline
   54& 3.5 & 60 &&58 & 6.5 & 69 && 81 & 5.0  & 88 & 12.2 & 34.4\\
   \hline\hline
  \end{tabular}
 \end{center}
\end{table}

The overlap between the ground state and the Bloch-Brink wave function amounts to approximately
50 to 60\% which qualitatively consistent with the Hartree-Fock results discussed in the previous
section. The estimated inter-cluster distance is 3.5 fm which is larger than that of Hartree-Fock
result. This owes to the angular momentum projection which lowers energy of the deformed
configuration and induces larger deformation of the ground state. The $0^+_4$ and $1^-_1$ states
have larger overlap with the Bloch-Brink wave function and the larger inter-cluster distances than
the ground state. This is apparently due to the excitation of the inter-cluster motion, which
enlarges the inter-cluster distance and reduces the cluster distortion. 

The calculated IS monopole and dipole transition matrix from the ground state to the $0^+_4$ and
$1^-_1$ states are also listed in Tab. \ref{tab:Ne20ism}. It is evident that those excited cluster
states have large transition matrix that are a few times larger than the Weisskopf estimates.
This result suggests that both of the positive- and negative-parity cluster states can be strongly
generated by the IS monopole and dipole transitions which will be good signature to identify the
asymmetric clustering. 

To summarize so far, I have discussed the dual character of shell and cluster, and IS monopole
and dipole transitions which activate the degrees-of-freedom of cluster excitation. By the
analysis of the Hartree-Fock wave functions, we have seen that the ground states of 
$^{8}{\rm Be}$, $^{16}{\rm O}$ and $^{20}{\rm Ne}$ have large overlap with Bloch-Brink wave
functions. This fact means that the ground state has dual character of shell and cluster, and can
be reasonably expressed by both of shell and cluster model wave functions. This dual character
does not necessarily mean that the ground state has cluster structure, but means that the
degrees-of-freedom of cluster excitation is embedded in the ground state. Therefore, both of the
single-particle and cluster excitations are possible from the ground state. 

As  triggers to induce the cluster excitation, I have discussed IS monopole and dipole
transitions of $^{20}{\rm Ne}$. I have shown that those operators bring about the nodal and
angular excitations of the inter-cluster motion. By an analytic calculation, I have shown that the
transition matrices from the ground state to the cluster states are comparable with the Weisskopf
estimates, even if the ground state has an ideal shell structure.  Furthermore, by using the
Bloch-Brink wave function, it was shown that the IS dipole transition matrix is considerably
amplified if the ground state has cluster structure. Finally, quantitative evaluation of the
transition matrix was made by using AMD. As a result, it is confirmed that IS monopole and
dipole transition matrices are sensitive to the cluster states. Therefore, it looks promising to
search for unknown cluster states in heavier mass system such as $^{24}{\rm Mg}$ \cite{kw13} and  
$^{28}{\rm Si}$ \cite{it13} by using IS monopole and dipole transitions as probe.

\section*{Linear-chains in neutron-rich carbon isotopes}
\begin{figure}[h]
 \begin{center}
  \includegraphics[width=0.9\hsize]{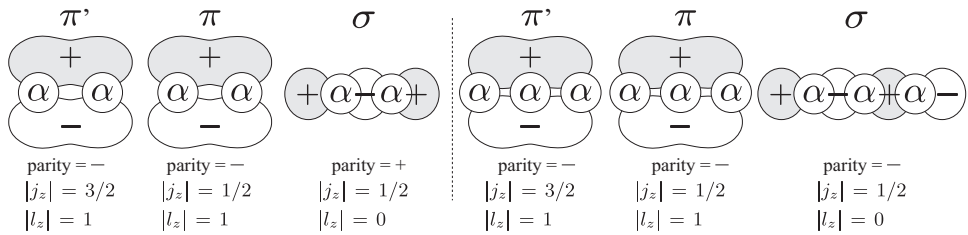}
  \caption{Left: molecular orbits in Be isotopes that have 2$\alpha$ cluster core. Right:
  An analogy of molecular orbits in 3$\alpha$ systems}\label{fig:ill}   
 \end{center}
\end{figure}

In the study of neutron-rich nuclei, it was realized that the excess neutrons can induce or stabilize
exotic cluster structure. The extreme but unstable 2$\alpha$ cluster structure of $^{8}{\rm Be}$
is stabilized by adding one or two neutrons ($^{9}{\rm Be}$ and $^{10}{\rm Be}$ ). Furthermore, in
more neutron-rich nuclei such as $^{11}{\rm Be}$ and $^{12}{\rm Be}$, the excess neutrons enhance
2$\alpha$ clustering and enlarge the inter-cluster distance. As a result, neutron-rich Be isotopes
have a novel type of cluster structure in which the pronounced cluster core is surrounded and
sustained by excess neutrons like a molecule
\cite{sey81,oer96,ita00,des02,eny95,eny03,oer97,oer06}. It 
was found a special class of neutron single-particle orbits called ``molecular orbits''
\cite{sey81,oer96,ita00} is formed around the 2$\alpha$ cluster core, and play an essential role for
stabilizing and inducing the clustering. Two kinds of molecular orbits so-called $\pi$- and
$\sigma$-orbits shown in Fig. \ref{fig:ill} are important. Roughly speaking, $\pi$-orbit plays a
glue-like role and reduce the inter-cluster distance, while $\sigma$-orbit enhances the clustering
and enlarges the inter-cluster distance. It was shown that the combinations of the excess neutrons
occupying $\pi$- and $\sigma$-orbits reasonably explain the observed excitation spectra of Be
isotopes. 

This finding initiated the search for the linear-chain state of neutron-rich Carbon isotopes which
is composed of  linearly aligned three $\alpha$ particles with surrounding neutrons
\cite{oer97,oer06,ita01}. The linear-chain configuration of 3$\alpha$ particle was originally
suggested by Morinaga \cite{mor56} to explain the structure of the Hoyle state of
$^{12}{\rm C}$. Later it was found that the Hoyle state is not a linear-chain state, but a dilute
gas-like 3$\alpha$ cluster state. Nowadays, it is believed that the linear-chain structure of
3$\alpha$ particles is unstable against the bending motion (a perturbation which deviates from the
linear alignment of $\alpha$ particles). Therefore, an extra mechanism is needed to realize  the
linear chain configuration. One of the strong candidates is the addition of the  valence neutrons
occupying the 
molecular orbit. It was shown that if the neutrons occupy the so-called $\sigma$-orbit, a deformed
configuration of the core with an elongated shape is favored. Eventually, it was shown that the
linear-chain configuration having $\sigma^2\pi^2$ configuration is possibly stabilized in 
$^{16}{\rm C}$ \cite{ita01}. Motivated by this results, there are many theoretical studies on
linear-chain of Carbon isotopes \cite{mar10,suh11,kim14,zha15,bab16}. One major contribution is the
studies based on the Hartree-Fock calculations \cite{mar10,zha15}. As a result of early stage
research, I introduce the work  by Maruhn {\it et al.} \cite{mar10}, and the recent situation is
covered by N. Itagaki. Another contribution is from AMD \cite{suh11,kim14,bab16}, and I introduce a
recent application. In both cases, the valence neutron configuration and stability of the linear
chain are main concern, and I focus on $^{16}{\rm C}$ in the following.

\subsection*{A Hartree-Fock study}
In Ref. \cite{mar10}, Maruhn and his collaborators performed the Hartree-Fock calculations to
investigate the valence neutron configuration and stability of linear chain of $^{16}{\rm C}$. For 
this purpose, they started the iterative energy minimization from the initial states that have an 
ideal linear-chain configurations, that are constructed as follows. 
\begin{itemize}
 \item Gaussian wave functions with spin and isospin saturated quartets are placed at z = 0 and
       $z= \pm d$ fm, to represent the linearly aligned 3$\alpha$ clusters. The inter-cluster
       distance $d=3$ fm and the size of $\alpha$ cluster $\sigma=1.8$ fm were chosen as the
       initial condition.  
 \item Neutrons occupying the $\sigma$-orbit were represented by  deformed Gaussians (3 times 
      broader in the $z$-direction) multiplied by $z(z-d)(z+d)$ to describe the nodes of
      $\sigma$-orbit. 
 \item Neutrons in $\pi$-orbit were built by  similar deformed Gaussians but multiplied by
      $x \pm iy$. By changing the combination with the direction of spin, two different
      $\pi$-orbits denoted by $\pi'$ and $\pi$ are introduced. The former has $j_z=\pm 3/2$ and
      the spin-orbit interaction acts attractive, while the latter has $j_z=\pm 1/2$ and the
      repulsive spin-orbit interaction.
 \item Another neutron orbit denoted by $\delta$-orbit was also introduced and described by 
      the deformed  Gaussian multiplied by $z(x \pm iy)$.
\end{itemize}
By combining those $\sigma,\pi',\pi$ and $\delta$ orbits with 3$\alpha$ linear chain, they
prepared initial states having various linear chain configurations. If a certain linear-chain 
configuration is (meta)stable, it stays in a linear-chain configuration during a certain period of
the iteration and eventually decays into the ground state. On the other hand, if a linear-chain  
configuration is unstable, it will decay into other configurations quickly. 

\begin{figure}[h]
 \begin{center}
  \includegraphics[width=0.4\hsize]{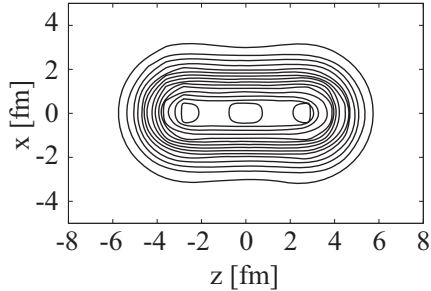}
  \caption{Total density distribution for the $\pi'^2\delta^2$ configuration of 
  $^{16}{\rm C}$. This figure is taken from Ref. \cite{mar10}.}\label{fig:C16dens}   
 \end{center}
\end{figure}

The obtained results were interesting and unexpected. They found that the four configurations
denoted by $\pi'^2\pi^2$, $\pi'^2\delta^2$, $\pi'^2\delta\pi$,  and $\pi'^2\sigma\pi$ are stable,
but other configurations decays into non-cluster states or populate an additional new linear-chain 
configuration denoted by $\pi'^2\delta\pi''$. The properties of those molecular orbits are
summarized in Tab. \ref{tab:mol0}.
\begin{table}[h]
 \begin{center}
  \caption{The parity, $|j_z|$ and number of nodes along $z$-axis $n_z$ of valence neutron orbits
  denoted by $\pi',\pi,\delta$ and   $\pi''$.   This table is reconstructed from
  Ref. \cite{mar10}.}\label{tab:mol0}   
  \begin{tabular}{cccc}
   \hline\hline
   type&parity&$|j_z|$&$n_z$\\
   \hline
   $\pi'$&$-$&3/2&0\\
   $\pi$&$-$&1/2&0\\
   $\sigma$&$-$&1/2&3\\
   $\delta$&$+$&3/2&1\\
   $\pi''$&$-$&1/2&0\\
   \hline\hline
  \end{tabular}
 \end{center}
\end{table}
Those quantum numbers for $\pi', \pi$ and $\sigma$ orbits agree with those discussed in the
cluster models, while the $\delta$ and $\pi''$ orbits have not been considered in the preceding
studies. All of those linear chain configurations appeared
approximately 15 to 20 MeV in excitation energies as listed in Tab. \ref{tab:lin0}. 
\begin{table}[h]
 \begin{center}
  \caption{Excitation energies of  linear chain configurations of $^{16}{\rm C}$ obtained by
  different Skyrme force parameters. This table is taken from Ref. \cite{mar10}.}\label{tab:lin0}  
  \begin{tabular}{ccccccc}
   \hline\hline
   Force&$\pi'^2\pi^2$&$\pi'^2\delta^2$&$\pi'^2\delta\pi$& $\pi'^2\sigma\pi$ &
   $\pi'^2\delta\pi''$\\
   \hline
   SkI3 & 14.5 & 19.5 & 17.0 & 17.5 & 19.1 \\
   SkI4 & 15.7 & 19.9 & 17.6 & 18.0 & 19.7\\
   Sly6 & 15.4 & 18.9 & 17.0 & 17.3 & 19.0\\
   SkM* & 16.4 & 17.5 & 16.9 & 17.0 & 19.7\\
   \hline\hline
  \end{tabular}
 \end{center}
\end{table}
The quadrupole
deformation parameter $\beta$ of those meta-stable linear chains are listed in Tab. \ref{tab:lin1}. 
\begin{table}[h]
 \begin{center}
  \caption{Quadrupole deformation parameter $\beta$ of the ground state and linear chain
  configurations of $^{16}{\rm C}$ obtained by different Skyrme force parameters.
  This table is taken from Ref. \cite{mar10}.}\label{tab:lin1} 
  \begin{tabular}{ccccccc}
   \hline\hline
   Force&g.s.&$\pi'^2\pi^2$&$\pi'^2\delta^2$&$\pi'^2\delta\pi$& $\pi'^2\sigma\pi$ &
   $\pi'^2\delta\pi''$\\
   \hline
   SkI3 & 0.34 & 0.69 & 0.82 & 0.76 & 0.76 & 0.88\\
   SkI4 & 0.33 & 0.68 & 0.80 & 0.75 & 0.74 & 0.86\\
   Sly6 & 0.32 & 0.68 & 0.81 & 0.75 & 0.75 & 0.87\\
   SkM* & 0.28 & 0.66 & 0.79 & 0.73 & 0.73 & 0.85\\
   \hline\hline
  \end{tabular}
 \end{center}
\end{table}
From this table, we can confirm that all of those meta-stable configurations have very stretched
structure comparable with the hyperdeformation in which the ratio of the longest and shortest axis
of nucleus reaches to 3:1. The density distribution of $\pi'^2\delta^2$ configuration is shown in
Fig. \ref{fig:C16dens} an an example. 
They also analyzed the nucleon configurations. They
classified nucleon single particle orbits into two groups; (1) the twelve orbits that constitute
the 3$\alpha$ cluster core, and (2) the orbits of four valence neutrons. From the twelve orbits of
(1) they reconstructed a Slater determinant corresponding to the 3$\alpha$ cluster core and made a
cluster analysis which was already explained in the first part of this contribution. The results
of analysis is given in Tab. \ref{tab:lin2}. 
\begin{table}[h]
 \begin{center}
  \caption{$\alpha$ cluster analysis of linear chain states of $^{16}{\rm C}$. The cluster wave
  function is composed of 3$\alpha$ particles with radius $\sigma$ positioned at $z=0,\pm R$. 
  This result is obtained by the SkI3 parameter set. The data is taken from
  Ref. \cite{mar10}.}\label{tab:lin2}  
  \begin{tabular}{cccccc}
   \hline\hline
   &$\pi'^2\pi^2$&$\pi'^2\delta^2$&$\pi'^2\delta\pi$& $\pi'^2\sigma\pi$ &
   $\pi'^2\delta\pi''$\\
   \hline
   $\mathcal{O}_{BB}(\%)$ & 63 & 65 & 64 & 64 & 66 \\
   $R$ & 2.44 & 2.68 & 2.56 & 2.56 & 2.80\\
   $\sigma$ & 1.70 & 1.67 & 1.68 & 1.68 & 1.68 \\
   \hline\hline
  \end{tabular}
 \end{center}
\end{table}
Compared to the result  of $^{8}{\rm Be}$ given in Tab. \ref{tab:Be8}, it is very interesting to
see that the size of $\alpha$ particles is close to those of free $\alpha$ particles and
in $^{8}{\rm Be}$. Furthermore, the inter-cluster distance is comparable or even larger than the
$^{8}{\rm Be}$. From this table, we can conjecture that the $\pi$, $\pi'$ orbits tend to reduce
the inter-cluster distance, while the $\sigma$ and $\delta$ orbits enlarge it. 

Thus, the Hartree-Fock calculation showed the possible existence of the linear-chain states in
$^{16}{\rm C}$. It is surprising that the 3$\alpha$ cluster core having large overlap with cluster 
wave function was obtained without a priori assumption. The results of  Hartree-Fock analysis may
be summarized as follows. 
\begin{enumerate}
 \item Five different kinds of linear chain configurations are concluded as meta-stable local
       minimum state. All of those states appear around 15 to 20 MeV in excitation energies, which
       are much smaller than those predicted by the cluster model.
 \item All of those states have 3$\alpha$ cluster cores which have large overlap with 3$\alpha$
       Bloch-Brink wave function with linear configuration, whose parameters are close to those of
       $^{8}{\rm Be}$. This result confirms the formation of 3$\alpha$ core with linear
       configuration. 
 \item In addition to the $\pi', \pi$ and $\sigma$ orbits, $\delta$ and $\pi''$ orbits are found,
       and they possibly stabilize the linear chain.  On the other hand, the $\sigma^2\pi'^2$
       configuration which was suggested as the most stable configuration against the bending
       motion \cite{ita01} was not obtained. 
\end{enumerate}

\subsection*{An AMD study}
In addition to the Hartree-Fock analysis, several studies based on antisymmetrized molecular
dynamics have also been performed. Here I introduce a resent result on $^{16}{\rm C}$ given in
Ref. \cite{kim14}. In this study, an AMD calculation was performed to search for the linear-chain
state in $^{16}{\rm C}$. Advantages of the AMD calculation are in the projections of $J^\pi$, and
the generator coordinate method. By $J^\pi$ projection, the rotational energy is subtracted and we
can obtain more precise excitation energy. The GCM takes care of the orthogonality condition of
the linear-chain states to other states having smaller excitation energies, which potentially
increases the stability of the linear chain state. However, because the single particle wave
packets are limited to the Gaussian form, the description of the molecular orbits is not as good as
that of Hartree-Fock. Keeping those points in mind, we refer to the AMD results.
\begin{figure}[h]
 \begin{center}
  \includegraphics[width=0.6\hsize]{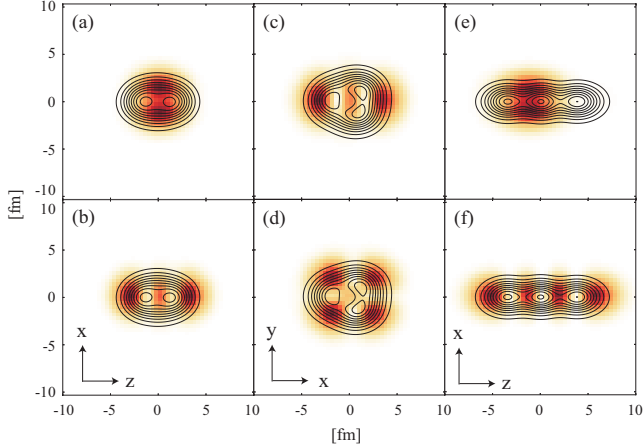}
 \caption{ The density distribution of the ground (a)(b), triangular (c)(d) and linear-chain
 (e)(f)  configurations. The contour lines show the proton density distributions and
 are common to the upper and lower panels. The color plots show the valence neutron orbits.
 The lower panels show the most weakly bound two neutrons, while the upper panel
 show the other two valence neutrons. This figure is taken from Ref. \cite{kim16}.} 
 \label{fig:C16density}
 \end{center}
\end{figure}
 
We first performed the constraint energy minimization to search for the strongly deformed minima. 
As a result, we obtained the global energy minimum corresponding to the ground state and two
local minima which have cluster structure. The density distributions of the core ($^{12}{\rm C}$)
and those of four valence neutrons are shown in Fig. \ref{fig:C16density}, and the properties of
valence neutron orbits are listed in Tab. \ref{tab:spo}.

\begin{table}[h]
\caption{The properties of valence neutron orbits shown in Fig. \ref{fig:C16density} which are
 occupied by two neutrons. Each column  show the single particle energy $\varepsilon$ in MeV, the
 amount of the positive-parity component  $p^+$ and the  angular momenta. This table is taken from
 Ref. \cite{kim16}}
\label{tab:spo}
\begin{center}
  \begin{tabular}{ccccccc} 
   \hline\hline
	orbit & $\varepsilon $ & $p^+$ & $j$ & $|j_{z}|$ & $l$ & $|l_{z}|$ \\ \hline
	(a) & $-8.24$ & 0.00 & 0.75 & 0.51 & 1.05 & 0.97 \\
	(b) & $-5.23$ & 0.99 & 2.21 & 0.51 & 1.80 & 0.38 \\ \hline
	(c) & $-5.74$ & 0.99 & 2.31 & 1.96 & 1.93 & 1.63 \\
	(d) & $-3.29$ & 0.98 & 2.33 & 1.88 & 2.07 & 1.83 \\ \hline
	(e) & $-5.32$ & 0.13 & 2.09 & 1.49 & 1.72 & 0.99 \\
	(f) & $-4.18$ & 0.03 & 2.89 & 0.53 & 2.72 & 0.18 \\ 
   \hline\hline
  \end{tabular}
\end{center}
\end{table}

As seen in its proton and valence neutron density distribution shown in Fig.\ref{fig:C16density}
(a) and (b), the ground state has no pronounced clustering. Four valence neutrons have an
approximate $(0p_{1/2})^2(0d_{5/2})^2$  configuration that is confirmed from the properties of
neutron single particle orbits listed  in  Tab. \ref{tab:spo}. The deviation from the spherical
$p_{1/2}$ and $d_{5/2}$ orbits is due to prolate deformation of this state.  
As deformation increases, valence neutron configuration is changed and  induces 3$\alpha$
clustering. A triaxially deformed 3$\alpha$ cluster configuration shown in
Fig. \ref{fig:C16density} (c) and (d) appears. This configuration has 3$\alpha$ cluster core of an
approximate isosceles triangular configuration with 3.2 fm long sides and 2.3 fm short side 
surrounded by valence neutrons occupying an approximate $(0d_{5/2})^4$ configuration.

Further increase of nuclear deformation realizes the linear-chain structure shown in
Fig. \ref{fig:C16density} (e) and (f).  As clearly seen in its density distributions, a
linearly aligned 3$\alpha$ cluster core is accompanied by four valence neutrons.
The interpretation of this configuration is given by the molecular
orbit picture. Namely, as confirmed from the properties listed in Tab. \ref{tab:spo}, the valence
neutron orbits are in good accordance with the  $\pi'$ and $\sigma$ orbits  found in the
Hartree-Fock result, and hence understood as $(\sigma)^2(\pi')^2$ configuration. 
\begin{figure*}[h]
 \begin{center}
 \includegraphics[width=0.9\hsize]{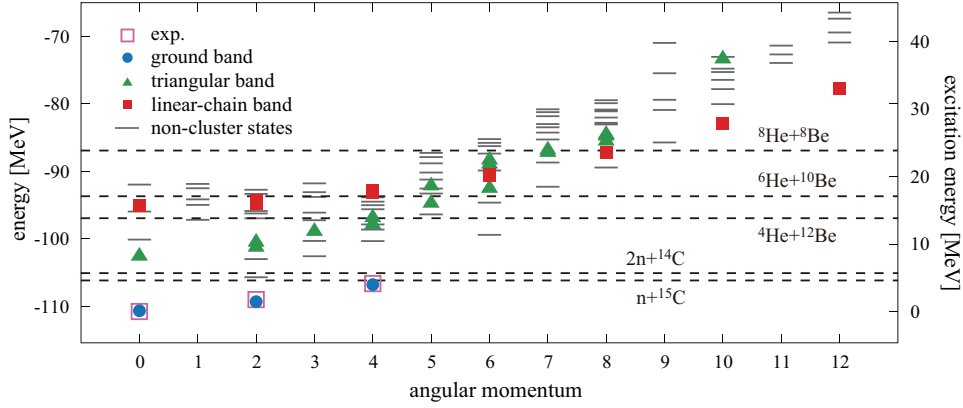}
 \caption{ The calculated and observed positive-parity energy levels of $^{16}$C up to
 $J^\pi=12^+$  states. Open boxes show the observed states with the definite spin-parity assignments, and
 other  symbols show the calculated result. This figure is taken from Ref. \cite{kim16}.}
 \label{fig:spectrum} 
 \end{center}
\end{figure*}

Figure \ref{fig:spectrum} shows the spectrum obtained by the angular momentum projection and GCM
calculation, where  the obtained states are classified to the 'ground band', 'triangular band',
'linear-chain band' and other non-cluster states by referring their cluster overlap. The member
states of the ground band shown by circles are dominantly composed of the wave functions
with $(sd)^2$ configuration. The calculated energy of the ground band is reasonably described.
Owing to its triaxial deformed shape, the triangular configuration generates two rotational bands
built on the $0^+_2$ and $2^+_5$ states. The member states have large overlap with the basis wave
function shown in Fig. \ref{fig:C16density} (c)(d).

The  linear-chain configuration appears as a rotational band built on the $0^+_5$ state at 15.5 MeV,
that is close to the $^{4}$He+$^{12}$Be and $^{6}$He+$^{10}$Be threshold energies. The band head state
$0^+_5$ has the largest overlap with the basis wave function shown in Fig.\ref{fig:C16density}
(e)(f). The moment-of-inertia is estimated as $\hbar/2\Im=112$ keV. Naturally, as the 
angular momentum increases, the excitation energy of the linear-chain state is lowered relative to
other structures, and the $J^\pi=10^+$ member state at $E_x=27.8$ MeV becomes the yrast
state.  Thus, we predict the stable  linear-chain configuration with molecular-orbits whose
band-head energy is around $^{4}$He+$^{12}$Be and $^{6}$He+$^{10}$Be thresholds.

One of the  concerns about the linear-chain configuration is its stability against the bending
motion, and we confirm it by investigating the response to $\gamma$ deformation.  

\begin{figure}[h]
 \begin{center}
 \includegraphics[width=0.5\hsize]{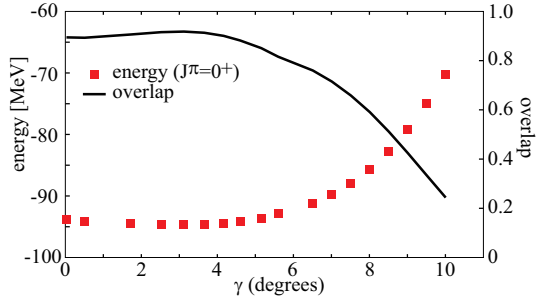}
 \caption{ The boxes show the energy of the linear-chain configuration with  $J^\pi=0^+$
 as function of quadrupole deformation parameter $\gamma$. The solid line shows the overlap
 between the linear-chain state ($0^+_5$ state) and the basis wave functions.
  This figure is taken from Ref. \cite{kim16}} 
 \label{fig:gamma} 
 \end{center}
\end{figure}

Starting from the linear-chain configuration shown in Fig. \ref{fig:C16density} (e)(f), we
gradually increased quadrupole deformation parameter $\gamma$ but kept $\beta$ constant to force
the bending motion. Thus obtained curve shown in Fig. \ref{fig:gamma} by squares  corresponds to the
energy surface against the bending motion. It is almost constant for small value of $\gamma$, but
rapidly increases for larger values of $\gamma$ indicating its stability. The solid line  in
Fig. \ref{fig:gamma} shows the overlap between the linear-chain $0^+_5$ state  and the  basis wave
function shown by squares that is defined as, 
\begin{align}
 O(\gamma) = |\langle\Psi(0^+_5)|\Phi(0^+;\gamma)\rangle|^2.
\end{align}
Here, $\Psi(0^+_5)$ and $\Phi(0^+;\gamma)$ denote the GCM wave function of the $0^+_5$ state and
the linear chain configuration with bending motion. The overlap  has its maximum at $\gamma=3.1$
degrees and falls off very quickly as $\gamma$ increases. Therefore the wave function of the 
linear-chain state is well confined within a region of small $\gamma$, and hence stable against
the bending motion. We conjecture that this stability of linear-chain configuration originates in
the orthogonality condition. Note that the energy of the linear-chain configuration is higher than
that of the triangular configuration. Because the linear-chain state must be orthogonal to the
triangular state, the bending motion is prevented by this orthogonality condition. 

Thus, both of the Hartree-Fock and AMD calculations predict the existence of the linear-chain
states of $^{16}{\rm C}$ around 15 MeV. However, as listed below, there are several disagreements
between those models that require further investigations. 
\begin{enumerate}
 \item The favored valence neutron configurations are quite different. The Hartree-Fock suggested
       five meta-stable configurations, but none of them was found in AMD calculation. On the
       other hand, the cluster model and AMD suggest the stability of $(\pi')^2(\sigma)^2$
       configuration, which was turned out unstable in Hartree-Fock.
 \item Nevertheless, the excitation energies of the linear chain in Hartree-Fock and AMD are
       similar  to each other. There could be some universal rule that determines the excitation
       energy. 
 \item The discussion about the stability of the linear chain still remains rather simple
       estimations. The theoretical investigation of the $\alpha$ and neutron decay widths should
       be made. They are also important to identify the linear chain from the observed data.
\end{enumerate}
Very recently, a couple of promising experimental data for linear chain configuration in for
$^{14}{\rm C}$ \cite{fre14,fri15,yam15,Tia16} and $^{16}{\rm C}$ \cite{koy15,del16} were reported by
several groups. We believe that further theoretical and experimental studies will reveal the
linear chain in neutron-rich Carbon isotopes in near future.

\section*{Summary}
In this contribution, I have discussed two topics. In the first part, the dual character of shell
and cluster suggested by Bayman-Bohr theorem is discussed. The Hartree-Fock analysis was very
helpful to understand the dual character. Owing to this dual character, both of the
single-particle and cluster excitations (and of course as well as the collective excitations) are
regarded as essential excitation modes of atomic nucleus. Indeed, by using $^{20}{\rm Ne}$ as an
example, I've demonstrated that IS monopole and dipole transitions indeed induce the cluster
excitation. Therefore, I expect that those transitions will be very powerful probe to search for
the cluster states. 

In the second half of the contribution, I discussed the linear-chain of the 3$\alpha$ particles in
neutron-rich Carbon isotopes. Realization of such one dimensional structure is a long dream in
nuclear physics, but found to be unrealistic in stable nucleus of $^{12}{\rm C}$. However, by the
development in the physics of unstable nuclei, people realized that the valence neutrons may
assist the cluster core to form the linear chain. Possible formation of such structure was
investigated by Hartree-Fock and AMD. It was surprising that Hartree-Fock suggests various kind of  
linear chains, and both of Hartree-Fock and AMD predict similar excitation energies. However,
there are several striking disagreements in their results, that must be resolved. With the help of
the increasing experimental data, I believe that the linear chain states will be identified in
near future.

\section*{Acknowledgment}
The author greatly thanks that the discussion with Dr. Maruhn and his collaborators is always
exciting, creative and productive. The topics discussed in this contribution is a part of the
results hatched from the discussions with Dr. He also acknowledges that part of the numerical
calculations were performed on the supercomputer at KEK and YITP. This work was supported by the
Grants-in-Aid for Scientific Research on Innovative Areas from MEXT (Grant No. 2404:24105008) and
JSPS KAKENHI Grant No. 25400240.

% figure
%\begin{figure} \begin{center}
%\includegraphics[width=10.0cm]{fig1.eps}    
%\caption{Caption of Fig.1} 
%\end{center} \end{figure}


\begin{thebibliography}{99}
% IEEE Style
 \bibitem{gam31} G. Gamow, {\it Constitution of Atomic Nuclei and Radioactivity}, 
	 (Oxford University Press, 1931).
 \bibitem{cha32} J. Chadwick, Nature {\bf 192}, 312 (1932).
 \bibitem{wil58} K. Wildermuth and Th. Kanellopoulos, Nucl. Phys. {\bf 7},  150 (1958).
 \bibitem{wil59} K. Wildermuth and Th. Kanellopoulos, Nucl. Phys. {\bf 9},  449 (1958/59).
 \bibitem{whe37} J. A. Wheeler, Phys. Rev. {\bf 52}, 1083 (1937).
 \bibitem{wil77} K. Wildermuth and Y. C. Tang, {\it A Unified Theory of the Nucleus} (Academic, New
	 York, 1977).
 \bibitem{bri66} D. M. Brink, Proc. Int. School of Physics Enrico Fermi, Course 36, Varenna,
	 ed. C. Bloch (Academic Press, New York, 1966).
 \bibitem{bec10} C. Beck (Ed.), {\it Clusters in Nuclei Vol. 1}, Lecture Notes in Physics,
	 vol. 818, (Springer, Berlin, Heidelberg, 2010).
 \bibitem{bec12} C. Beck (Ed.), {\it Clusters in Nuclei Vol. 2}, Lecture Notes in Physics,
	 vol. 848, (Springer, Berlin, Heidelberg, 2012).
 \bibitem{bec14} C. Beck (Ed.), {\it Clusters in Nuclei Vol. 3}, Lecture Notes in Physics,
	 vol. 875, (Springer, Berlin, Heidelberg, 2014).
 \bibitem{bay58} B. F. Bayman, and A. Bohr, Nucl. Phys. {\bf 9}, 596 (1958/1959).
 \bibitem{ell58} J. P. Elliott, Proc. R. Soc. London A {\bf 245}, 562 (1958).
 \bibitem{hor68} H. Horiuchi, and K. Ikeda, Prog. Theor. Phys. {\bf 40}, 277 (1968).
 \bibitem{ueg77} E. Uegaki, S. Okabe, Y. Abe, and H. Tanaka,
	 Prog. Theor. Phys. {\bf 57}, 1262 (1977).
 \bibitem{kam81} M. Kamimura, Nucl. Phys. A {351}, 456 (1981).
 \bibitem{des87} P. Descouvemont, and D. Baye, Phys. Rev. C {\bf 36}, 54 (1987).
 \bibitem{kan98} Y. Kanada-En'yo, Phys. Rev. Lett. {\bf 81}, 5291 (1998).
 \bibitem{toh01} A. Tohsaki, H. Horiuchi, P. Schuck, and G. R\"opke, 
	 Phys. Rev. Lett. {\bf 87}, 192501 (2001).
 \bibitem{nef04} T. Neff, and H. Feldmeier, Nucl. Phys. A {\bf 738}, 357 (2004).
 \bibitem{yam08-1} T. Yamada, H. Horiuchi, K. Ikeda, Y. Funaki and A. Tohsaki, Jour. of
	 Phys. Conf. Ser. {\bf 111}, 012008 (2008). 
 \bibitem{fun06} Y. Funaki, A. Tohsaki, H. Horiuchi, P. Schuck, and G. R\"opke,
		 Eur. Phys. J A {\bf 28}, 259 (2006). 
 \bibitem{fun15} Y. Funaki, H. Horiuchi and A. Tohsaki, Prog. Part. Nucl. Phys. {\bf 82}, 78 (2015).
 \bibitem{suz87} Y. Suzuki, Nucl. Phys. A {\bf 470}, 119 (1987).
 \bibitem{suz89} Y. Suzuki, and S. Hara, Phys. Rev. C {\bf 39}, 658 (1989).
 \bibitem{yam08} T. Yamada, Y. Funaki, H. Horiuchi, K. Ikeda, and A. Tohsaki, 
	 Prog. Theor. Phys. {\bf 120}, 1139 (2008).
 \bibitem{hor10} H. Horiuchi, J. Phys. Conf. Ser. {\bf 569}, 012001 (2014). 
 \bibitem{kim16} Y. Chiba, M. Kimura and Y. Taniguchi, Phys.Rev. C {\bf 93}, 034319 (2016).
 \bibitem{sey81} M. Seya, M. Kohno, and S. Nagata, Prog. Theor. Phys. {\bf 65}, 204 (1981).
 \bibitem{oer96} W. von Oertzen, Z. Phys. A {\bf 354}, 37 (1996);  {\it ibid} {\bf 357}, 355 (1997).
 \bibitem{ita00} N. Itagaki and S. Okabe, Phys. Rev. C {\bf 61}, 044306 (2000).
 \bibitem{des02} P. Descouvemont, Nucl. Phys. A {\bf 699}, 463 (2002).
 \bibitem{eny95} Y. Kanada-En'yo, H. Horiuchi, and A. Ono, Phys. Rev. C {\bf 52}, 628 (1995).
 \bibitem{eny03} Y. Kanada-En'yo and H. Horiuchi, Phys. Rev. C {\bf 68}, 014319 (2003).
 \bibitem{oer97} W. von Oertzen, Z. Physik A {\bf 357}, 355 (1997).
 \bibitem{oer06} W. von Oertzena, M. Freer, and Y. Kanada-En'yo, Phys. Rep. {\bf 432}, 43 (2006).
 \bibitem{ita01} N. Itagaki, S. Okabe, K. Ikeda, and I. Tanihata, Phys. Rev. C {\bf 64}, 014301
	 (2001).
 \bibitem{mar10} J. A. Maruhn, N. Loebl, N. Itagaki, and M. Kimura, Nucl. Phys. A {\bf 833}, 1
	(2010).
 \bibitem{suh11} T. Suhara, Y. Kanada-Enyo,  Phys. Rev. C {\bf 84}, 024328 (2011).
 \bibitem{kim14} T. Baba, Y. Chiba, and M. Kimura, Phys. Rev. C {\bf 90}, 064319 (2014). 
 \bibitem{zha15} P. W. Zhao, N. Itagaki, and J. Meng, Phys. Rev. Lett. {\bf}115 , 022501 (2015). 
 \bibitem{bab16} T. Baba,  and M. Kimura, Phys. Rev. C {\bf 94}, 044303 (2016). 
 \bibitem{fre14} M. Freer {\it et al}., Phys. Rev. C {\bf 90}, 054324 (2014). 
 \bibitem{koy15} S. Koyama {\it et al.}, private communication.
 \bibitem{del16} D. Dell'Aquila {\it et al.}, Phys. Rev. C {\bf 93}, 024611 (2016).
 \bibitem{fri15} A. Frisch {\it et al.}, Phys. Rev. C {\bf 93}, 014321 (2016).
 \bibitem{yam15} H. Yamaguchi {\it et al.}, arXiv:1610.06296.
 \bibitem{Tia16} Z. Y. Tian {\it et al.}, Chi. Phys. C {\bf 40}, 111001 (2016).
 \bibitem{mar06} J. A. Maruhn, M. Kimura, S. Schramm, P.-G. Reinhard, H. Horiuchi, and A. Tohsaki
	 Phys. Rev. C {\bf 74}, 044311 (2006). 
 \bibitem{wir00} R. B. Wiringa, S. C. Pieper, J. Carlson, and V. R. Pandharipande, Phys. Rev. C
	 {\bf 62}, 014001 (2000). 
 \bibitem{hor77} H. Horiuchi, Prog. Theor. Phys. Suppl. {\bf 62}, 90 (1977).
 \bibitem{zha94} J. Zhang, W. D. M. Rae, and A. C. Merchant, Nucl. Phys. A {\bf 575}, 61 (1994).
 \bibitem{mat75} T. Matsuse, M. Kamimura, and Y. Fukushima, Prog. Theor. Phys. {\bf 53}, 706
	 (1975). 
 \bibitem{buc75} B. Buck, C. B. Dover, and J. -P. Vary, Phys. Rev. C {\bf 11}, 1803 (1975).
 \bibitem{hee76} P. -H. Heenen, Nucl. Phys. A {\bf 272}, 399 (1976).
 \bibitem{fuj79} Y. Fujiwara, H. Horiuchi, and R. Tamagaki, Prog. Theor. Phys. {\bf 62}, 122
	 (1979).
 \bibitem{kru92} M. Kruglanski and D. Baye, Nucl. Phys. A {\bf 548}, 39 (1992).
 \bibitem{duf94} M. Dufour. P. Descouvemont, and D. Baye, Phys. Rev. C {\bf 50}, 795 (1994).
 \bibitem{bo13} B. Zhou, Y. Funaki, H. Horiuchi, Z. Ren, G. R\"opke, P. Schuck, A. Tohsaki,
	 C. Xu, and T. Yamada, Phys. Rev. Lett. {\bf 110}, 262501 (2013).
 \bibitem{zhou15} E. F. Zhou, J. M. Yao, Z. P. Li, J. Meng, P. Ring, Phys. Lett. B, in print.
 \bibitem{ang13} I. Angelia, and K. P. Marinovab, Atom. data and Nucl.Data Tables, {\bf 99}, 69
	 (2013).
 \bibitem{mart} J. A. Maruhn and W. Greiner, {\it Nuclear Models} (Springer, 1996).
 \bibitem{amd1} Y. Kanada-En'yo, M. Kimura and H. Horiuchi, C. R. Physique {\bf 4},  497 (2003).
 \bibitem{kim04} M. Kimura Phys. Rev. C {\bf 69}, 044319 (2004).
 \bibitem{amd2} Y. Kanada-En'yo, M. Kimura and A. Ono, PTEP {\bf 2012},  01A202 (2012).
 \bibitem{gog91} J. F. Berger, M. Girod, and D. Gogny, Comput. Phys. Comm. {\bf 63} (1991) 365.
 \bibitem{kim12} M. Kimura, R. Yoshida, and M. Isaka,  Prog. Theor. Phys. {\bf 127}, 287 (2012).
 \bibitem{hill53} D. L. Hill and J. A. Wheeler, Phys. Rev. {\bf 89}, 112 (1953).
 \bibitem{hill57} J. J. Griffin and J. A. Vfheeler, Phys. Rev. {\bf 108}, 311 (1957).
 \bibitem{kw13} T. Kawabata {\it et al.}, Jour. Phys. Conf. Ser. {\bf 436}, 012009 (2013).
 \bibitem{it13} M. Itoh {\it et al.}, Phys. Rev. C {\bf 88}, 064313 (2013). % 28Si
 \bibitem{mor56} H. Morinaga,  Phys. Rev. {\bf 101}, 254 (1956). 
\end{thebibliography}
\end{document}